\documentclass[aps,prb,twocolumn,10pt]{revtex4-1}
\usepackage{ifpdf}
\usepackage{amsmath}
\usepackage{graphicx}
\usepackage{epstopdf}
\usepackage{mathtools}
\usepackage{tikz}
\usepackage{hyperref}
\usepackage{color}
\begin{document}
\newcommand{\dkc}[1]{\textcolor{cyan}{[\textit{David}: #1]}}
\newcommand{\as}[1]{\textcolor{blue}{[\textit{Abhishek}: #1]}}
\newcommand{\nky}[1]{\textcolor{red}{[\textit{Nahom}: #1]}}
\title{Application of Generalized Periodic Anderson Hamiltonians to the Superconducting Nickelates}
\author{Abhishek Som}
\author{Nahom K. Yirga}
\author{David K. Campbell}
\affiliation{Department of Physics, Boston University, Boston, Massachusetts 02215, USA}
\date{\today}

\date{\today}
\begin{abstract}

We study the extent to which a three-dimensional dispersing Periodic Anderson Model (PAM) can explain the emergence of novel superconductivity in the Infinite-Layer Nickelate compounds. By going beyond frequently used 2D models, the 3D dispersing PAM allows us to incorporate effects of finite out-of-plane hopping and orbital hybridization in describing these systems. Using an unbiased functional Renormalization Group (fRG) approach, we show that $d_{x^2-y^2}$ superconductivity arises in a series of 3D {\it {ab-initio}} models of the Nickelates ({\it {e.g.}}$\mathrm{RNiO_2}$), where R is a rare earth element. We the study the impact of going beyond the Ni-d orbital by including the R-$d_{z^2}$ and the interstitial-$s$ as hybridizing conducting bands. We explore the dependence of the models on key parameters, including the local Hubbard coupling, doping and temperature. We find the hybridization with the interstitial-$s$ band driving a 3D $d_{z^2-r^2}$-type superconductivity while out of plane hopping primarily enhances an $s$-wave superconducting order.
\end{abstract}
\maketitle
\section{Introduction}
The recent discovery of superconductivity in $\mathrm{Sr}$-doped $\mathrm{NdNiO_2}$\cite{li2019superconductivity} has re-kindled interest in studies of single orbital models of high $T_c$ superconductivity due to the Sr-doping primarily affecting the occupation of the $3d_{x^2-y^2}$-Ni orbital\cite{held2022phase,kitatani2020nickelate}. However, DFT+DMFT studies of the Nickelates indicate a need to go beyond the single orbital Hubbard model, as the dominant $d_{x^2-y^2}$ orbital shows strong hybridization with the interstitial-s orbital in the Nd layer\cite{gu2020substantial}. The hybridization out of the $\mathrm{NiO_2}$ plane into the Nd layer also necessitates moving beyond the $\mathrm{NiO_2}$ planes to three dimensional models.

Previous studies of the impact of hybridization with a conduction band on a correlated system have been conducted primarily in the Periodic Anderson Model (PAM), which models a localized orbital with electronic correlations. Though PAM models were originally constructed to study {\it non-dispersing} f orbitals in heavy electron systems\cite{anderson1961localized}, they have been utilized in the study of the Cuprates, with the Cu-d and O-p serving as the localized and conducting bands, to explore Mott insulator transitions\cite{sordi2007metal}. Unlike the Cuprates, modeling the $\mathrm{NiO_2}$ planes of the Nickelates requires accounting for the rare earth 5d orbitals which contribute to the low-energy model and hybridize with $3d_{x^2-y^2}$-Ni band. The effects of hybridization with the rare earth bands requires a generalization of the PAM that allows for dispersion in the correlated orbital\cite{hepting2020electronic}. Recent studies of trends in the Nickelates have constructed a series of microscopic Hamiltonians designed to capture hybridization between the $d_{x^2-y^2}$ and the $5d_{z^2}$ orbitals in addition to the correlated $d_{x^2-y^2}$ band\cite{been2021electronic}. These models are of interest in view of the many different proposals for the symmetry of the superconducting gap observed in the Nickelates. The possibilities for the superconducting gap range from the standard $d_{x^2-y^2}$-wave to $d+s$-wave to $d_{xy}$ for the different rare earth metals R \cite{chow2022pairing,chen2022antiferromagnetic,wang2020distinct}. The picture of a singular superconducting order for the Nickelates is further complicated by studies of model Nickelate systems that show the superconducting order changing from $d_{x^2-y^2}$ to $s_{\pm}$ as a function of electronic correlations\cite{kreisel2022superconducting}. Thus, a complete treatment of the Nickelates requires accounting for electronic correlations, momentum dependent hybridizations and the response of the system to doping.

Momentum dependent hybridization within PAM models has previously been studied primarily in two dimensions. The interplay between local correlations and hybridization drives fluctuations in the local spins with electronic correlations enhancing moment formation, which can in turn be screened by the conduction electrons leading to the antiferromagnetic and ferromagnetic orders seen in these systems. Beyond simple screening, momentum dependent hybridization can introduce frustration, which can reduce or suppress antiferromagnetic correlations in the system. Previous studies of 2D PAMs with frustrated hybridization done with an eye to the Ce-115 systems have captured the emergence of $d_{x^2-y^2}$-type superconducting fluctuations as the degree of frustration is tuned\cite{wu2015d}. The spin fluctuations in the frustrated system still mediate pairing, with unconventional superconducting orders arising from a metallic state as the systems are doped. In three dimensions, momentum dependent hybridizations have been utilized in modeling the metallic rare earths and lead to the rapid screening of moments\cite{huscroft1999magnetic}. The effects of frustration and the presence of superconductivity remain unexplored in the 3D PAM, although screening due to hybridization with the 3D rare earth bands or the interstitial-s could be sufficient in explaining the observed
absence of antiferromagnetic order in the Nickelates. Previous studies of a locally hybridized 2D PAM via the dynamical vertex approximation have found criticality as a function of hybridization with the dispersive effects in the Nickelates presenting a new dimension to further enhance or suppress this criticality\cite{schafer2019quantum}.

The generalized PAM we consider here as a model for the Nickelates requires a dispersing correlated orbital and is equivalent to a Hubbard model with hybridization. Tuning the parameters of the model, we should be able to capture many of the phases seen in the PAM and Hubbard Hamiltonians. We utilize a decoupled variant of the functional Renormalization group (fRG) to construct the response of the generalized PAM in two and three dimensions. We explore the impact of the hybridization with the interstitial-$s$ and $d_{z^2}$ bands within the PAM framework and analyze the changes to the system's response for the case of a dispersive $d_{x^2-y^2}$ band. We find unconventional superconductivity of the $d_{z^2}$ and $d_{x^2-y^2}$-type in the three dimensional and two dimensional systems, respectively. The fRG allows us to explore the impact of doping the conduction band, the $s,d_{z^2}$-bands of the rare-earth layer, as well as the $d_{x^2-y^2}$ Ni-band.

The remaining sections of the manuscript are organized as follows. In Sec.\ref{modelPAMfrg} we introduce the hybridization and dispersions of the modified two- and three-dimensional PAM models and present the computational details of the fRG flow we employ to study these models. We discuss our results for the two-dimensional PAM systems in Sec.\ref{2DPAM}. In Sec.\ref{3DPAM} we present our results for the three-dimensional PAM models. In Sec. \ref{NickelSec} we introduce and construct flows for the {\it{ab-initio}} Hamiltonians proposed for the Nickelates\cite{been2021electronic}. We further analyze some simplified model Hamiltonians derived from the {\it{ab-initio}} models. We summarize our conclusions and present directions for future work in Sec.\ref{summary}

\section{Overview of Models and Sketch of the fRG Method}
\label{modelPAMfrg}

The Hamiltonian for the generalized periodic Anderson model (PAM) is given by
\begin{align}
\mathcal{H}&=-t_c\sum_{\langle ij\rangle\sigma}c_{i\sigma}^\dagger c_{j\sigma} -t_f\sum_{\langle ij\rangle\sigma}f_{i\sigma}^\dagger f_{j\sigma} +\nonumber\\
&\sum_{ij\sigma}V_{ij}(c_{i\sigma}^\dagger f_{j\sigma} + f_{j\sigma}^\dagger c_{i\sigma}) +U\sum_in_{i\uparrow}^fn_{i\downarrow}^f\nonumber\\&-\mu_c\sum_{i\sigma}n_{i\sigma}^c-\mu_f\sum_{i\sigma}n_{i\sigma}^f
\end{align}
where $c_{i\sigma}^\dagger$, $f_{i\sigma}^\dagger$ are operators for electrons at lattice site $i$ with spin $\sigma$ in the broad conduction and narrow (localized) bands respectively. The hopping amplitudes and chemical potentials of the bands are represented by $t_c$, $t_f$ and $\mu_c$, $\mu_f$. The electrons in the system interact primarily via the local Hubbard coupling ($U$) in the narrow (localized) band with the hopping between the localized and conduction bands ($V_{ij}$) presenting a channel that competes with and reduces the impact of the local repulsion. The basic PAM model is defined by a local, momentum independent hybridization (V) and a dispersionless localized band ($t_f=0$). Modeling systems beyond the Kondo insulators requires additional terms which at the single particle level requires allowing for extended hybridization ($V_{i,i+\delta}\neq 0$) and a broadening the localized orbital ($0<t_f<<t_c$). Such generalizations introduce frustration from the onsite and nearest neighbor hybridizations, which along with the broadened bands and Hubbard coupling determine the degree of localization of the f electrons. The interplay between hybridization and electronic correlation presents a rich playground with previous studies of PAM models finding quantum critical points as a function of the hybridization \cite{held2000similarities,held2000mott,schafer2019quantum}. Mapping the generalized PAM Hamiltonian to momentum space, we have
\begin{align}
\mathcal{H}=&\sum_{k\sigma}\xi_{k\sigma}^cc_{k\sigma}^\dagger c_{k\sigma} + \sum_{k\sigma}\xi_{k\sigma}^ff_{k\sigma}^\dagger f_{k\sigma} +\nonumber\\
&\sum_{k\sigma}V_k(c_{k\sigma}^\dagger f_{k\sigma} +f_{k\sigma}^\dagger c_{k\sigma})
\label{pamHamilt}
\end{align}
where $\xi_k^c$ is the dispersion of the conduction band, $\xi_k^f$ is that of the narrow (localized) orbital and $V_k$ represents a generalized hybridization. Most PAM systems are hybridized locally ($V_k=V_0$), but introducing a momentum dependence can help model those heavy fermion systems that show metallic or insulating states. In this work we primarily consider hybridizations of the form
\begin{align}
V_k=V_0+V_1(\cos(k_x)+\cos(k_y)+\cos(k_z))
\label{hybridizationV0V1}
\end{align}
\begin{align}
V_k=V_0+V_1\cos\left(\frac{k_z}{2}\right)(\cos(k_x)-\cos(k_y))
\label{hybridizationVIS}
\end{align}
\noindent
with the first expression corresponding to a band ($\xi_c$) hybridizing locally and to nearest neighboring sites on a cubic lattice while the second expression represents the overlap between the Ni-$d_{x^2-y^2}$ and an interstitial-$s$ band for the Nickelate systems. We note that the hybridizations can be projected to a 2D square lattice representative of the $NiO_2$ planes by averaging over the $k_z$ modes; for the interstitial-$s$ band, this simply leads to a reduced hybridization (see Sec.\ref{gPAM}).

Our goal is to study the responses generated by the hybridization and correlations in these interacting electron systems using the decoupled functional renormalization group outlined in previous works (fRG)\cite{yirga2021frequency}. For the PAM-type models, this begins by integrating out the non-interacting conduction electrons, which leads to a retarded momentum dependent potential, $\Delta_k(i\omega)$, for the f-electrons. The action associated with Hamiltonian defined in Eq.\ref{pamHamilt} is given by
\begin{align}
\mathcal{S}_\Lambda= &\sum_{\omega,k}(i\omega +R_\Lambda(\omega)-\Delta_k(\omega)-\xi_f-\Sigma_k^\Lambda(\omega))f_{\omega,k}^\dagger f_{\omega,k} +\nonumber\\
 &\sum_{k_1,k_2,k_3}\Gamma_{k_1,k_2,k_3,k_4}^\Lambda f_{k_1}^\dagger f_{k_2}^\dagger f_{k_3}f_{k_4}
\end{align}
where $\Gamma$ is the interaction vertex ($\Gamma^{\Lambda=st}=U$), $\Sigma_k^\Lambda(\omega)$ is the self energy and $R_\Lambda(\omega)$ is the regulator. The single particle potential from the hybridization with the conduction electrons is given by
\begin{align}
\Delta(\omega,k)=\frac{V_k^2}{i\omega-\xi_k^c}
\label{hybPot}
\end{align}
where the chemical potentials $\mu_c$ and $\mu_f$ have been subsumed in the corresponding dispersions. Starting at a scale ($\Lambda^{st}>>W$) where all free modes in $\mathcal{S}_{\Lambda}$ are frozen out, the fRG flow equations track the evolution of the vertices as modes are integrated out, scale by scale. The fRG flow of the vertices is constructed via the decoupling of two-body interactions as outlined in previous works\cite{lichtenstein2017high,schober2018truncated,yirga2021frequency}. The decoupling retains the singular pieces of the magnetic, charge and superconducting channels of the vertex while the auxiliary channels are expanded in a limited basis set. This truncation leads to a reduction of the full vertex from $\mathcal{O}(N_f^3N^3)$ to $\mathcal{O}(N_fN_\omega^2NN_k^2)$, with $N_f$ corresponding to the number of Matsubara frequencies retained, $N$ the number of momentum modes in the system and $N_\omega$, $N_k$ corresponding to the number of frequency and momentum basis functions used to approximate the auxiliary channels. Truncation of the auxiliary channels of the full vertex in each of the channels enables an unbiased and efficient treatment of system-wide fluctuations. From the flows for the self-energy ($\Sigma$) and the vertex ($\Gamma$), we study the quasi-particle weight
\begin{align}
\mathcal{Z}_k^{-1}=1-\partial_\omega\Sigma_k(\omega)\biggl{|}_{\omega=0}
\end{align}
the charge and spin response functions
\begin{align}
\chi_{ij}^{C,S}(\tau)=\langle (n_{i\uparrow}(\tau)\pm n_{i\downarrow}(\tau))(n_{i\uparrow}(0)\pm n_{i\downarrow}(0))\rangle
\end{align}
and the superconducting susceptibilities given by
\begin{align}
\chi^{SU}(q,\Omega)&=\sum_{\omega_1,k_1,\omega_2,k_2}f_{k_1}^{\mathcal{O}}f_{k_2}^{\mathcal{O}}\nonumber\\
&\langle f_{\omega_1,k_1,\uparrow}^\dagger f_{\Omega-\omega_1,q-k_1,\downarrow}^\dagger f_{\Omega-\omega_2,q-k_2,\downarrow}f_{\omega_2,k_2\uparrow}\rangle
\end{align}
where $f_k^{\mathcal{O}}$ is the momentum profile associated with the ordering, $\mathcal{O}$. For the models considered below the interactions drive momentum profiles beyond local ordering ($f_k^\mathcal{O}=1$) of the $d_{x^2-y^2}$ ($f_k^\mathcal{O}=\cos(k_x)-\cos(k_y)$), $d_{z^2-r^2}$ ($f_k^\mathcal{O}=2\cos(k_z)-\cos(k_x)-\cos(k_y)$) and extended-s types ($f_k^\mathcal{O}=\sum_i^D\cos(k_i)$). The responses observed are sensitive to the temperature at which the flow is constructed so we have set $\beta \sim 2N$ when studying dominant ordering tendencies and augmented these with studies of the temperature dependence of the system response when necessary.

\begin{figure}
  \centering
  \includegraphics[scale=0.295]{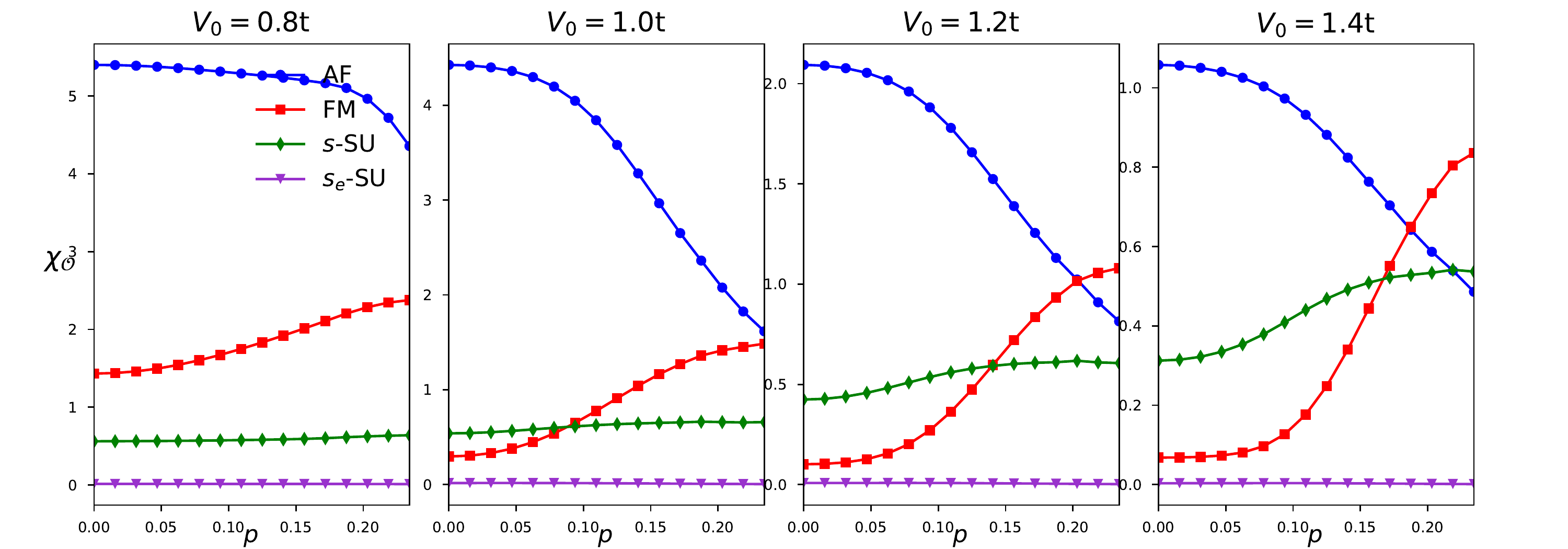}
  \includegraphics[scale=0.295]{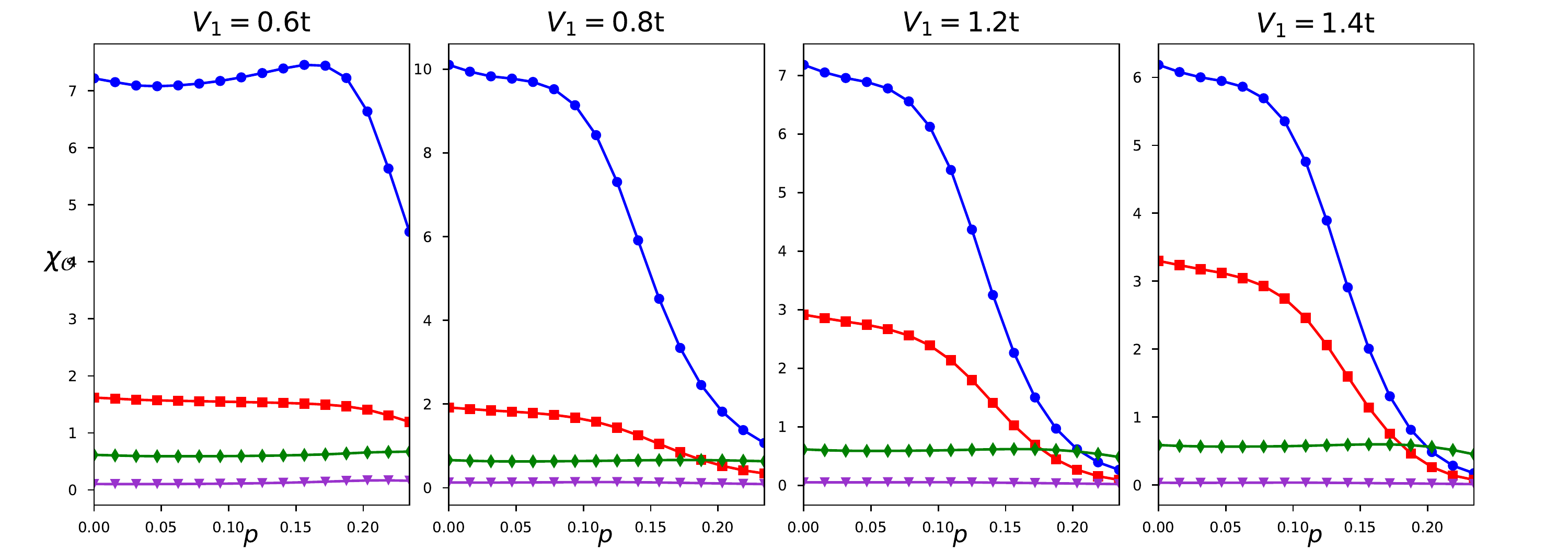}
  \includegraphics[scale=0.295]{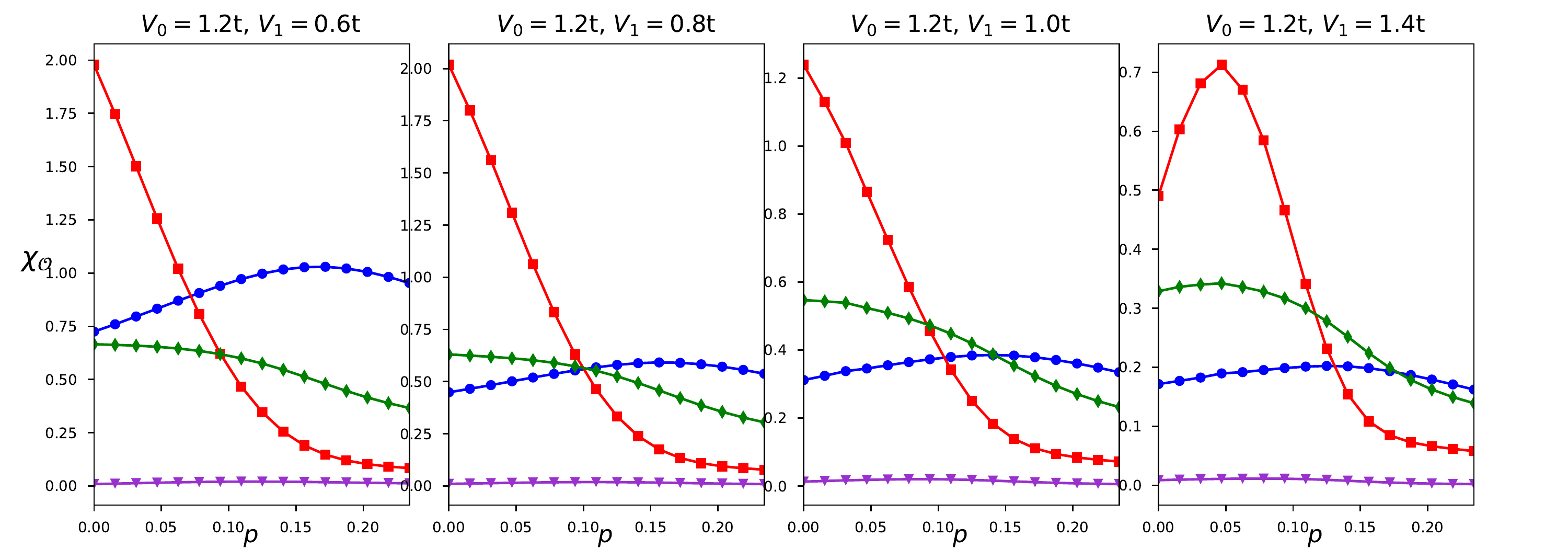}
  \caption{The antiferromagnetic, ferromagnetic, superconducting correlations ($s$, $d_{x^2-y^2}$) of the periodic Anderson model ($W_c=8t$) as a function of doping for different values of hybridization calculated via the 2-Loop fRG with $U=2t$, $N_\omega=4$, $N_k=2$ and $\beta=16$ for an $8\times 8$ lattice. Responses for a local hybridization of f-electrons ($V_0$), for a nearest neighbor hybridization ($V_1$) and the nearest neighbor hybridization ($V_1$) in the background of local hybridization ($V_0=1.2t$) are shown in the top, middle and bottom panels.}
  \label{frustrated_SU}
\end{figure}

\section{Two dimensional Periodic Anderson model}
\label{2DPAM}

The 2D PAM has served as a prototype model for describing the physics of insulating states in the heavy fermion compounds that can range from a band or Kondo insulator to an antiferromagnet\cite{anderson1961localized,vekic1995competition,zhang1988monte,yang2012emergent,schafer2019quantum}.  The competition between the hybridization ($V$), the electronic interaction ($U$) and bandwidth ($W_c$) has been well documented by Doniach and leads to states with ordered spins driven by $U$ that can be screened ($V_k$) by the conduction band ($W_c$)\cite{doniach1977kondo}.  The phase diagram of the model for local hybridization ($V_k=V)$ at half filling shows a transition from an antiferromanget to a Kondo insulator as a function of the hybridization\cite{vekic1995competition}. Models with an additional nearest neighbor hybridization have been used to model Ce-115 with Monte Carlo and dynamical cluster studies showing the emergence of $d_{x^2-y^2}$ superconductivity for parameter regimes relevant to the Ce system\cite{wu2015d}. The particular case of nearest neighbor hybridization shows a remarkable similarity to the Hubbard model with past studies finding a similar topology of phases and a Mott transition at finite Hubbard coupling\cite{held2000similarities,held2000mott}. Studies as to the effect of doping on the system remain minimal with constrained Monte Carlo studies finding a robust antiferromagnetic phase for a range of dopings without the incommensurate spin order observed in the Hubbard model\cite{bonvca1998effects,schulz1990incommensurate}. The model for the Ce-115 involved both a local and nearest neighbor hybridization with increased frustration ($V_0/V_1\sim 1$) expanding the superconducting domain. The interplay of doping with hybridization and correlation and the emergence of $d_{x^2-y^2}$ superconducting order is of relevance to the Nickelates, which show similar interactions and ordering tendencies.

\begin{figure}
    \centering
    \hspace*{-0.25cm}\includegraphics[scale=0.6]{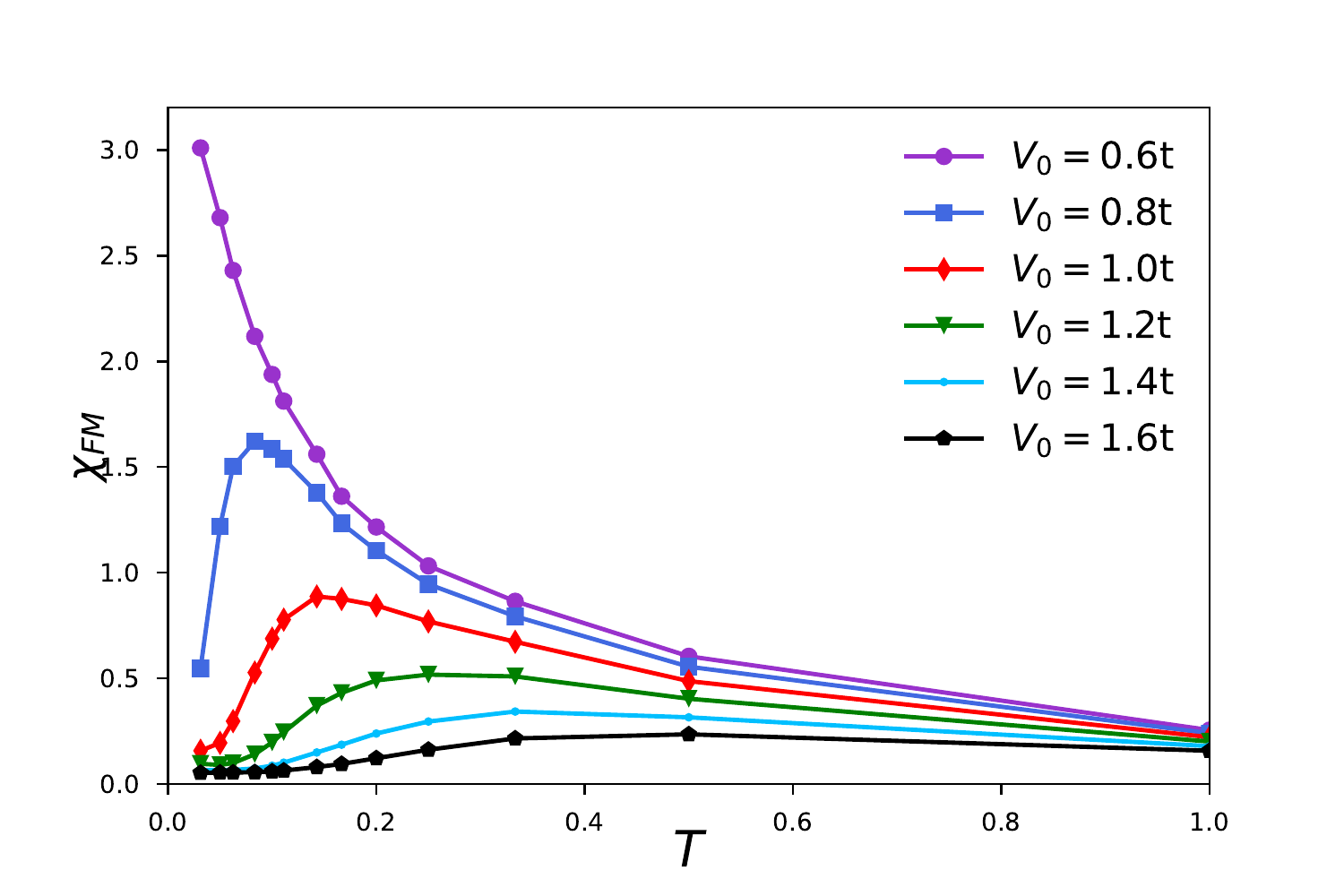}
    \caption{Ferromagnetic spin response of the 2D PAM, $U=2t$, $W_c=8t$, as a function of temperature for different values of local hybridization ($V_0$) to the conduction band ($W_c$). The spin response is screened with increasing hybridization.}
    \label{2DPAM_v0T}
\end{figure}

Calculation of the system response requires the construction of vertex flows that account for hybridization with local and nearest neighboring sites which requires specifying parameters for the decoupled fRG outlined above. The momentum profiles for the orders we wish to study are primarily local ($s,d$-type) and can be captured by setting the momentum resolution of each auxiliary channel to $N_k=2(3)$ which in two dimensions corresponds to 9 (13) basis functions, $(2N_k^x+1)(2N_k^y+1)$. Similarly the frequency resolution is set $N_\omega=4$ but can be enlarged to $N_\omega=6,8$ to account for the presence of two scales associated with the hybridization and dispersion. The frequency dependent single particle potential due to hybridization in the PAM can be sensitive to the choice of $\Delta T=\beta/N_\omega$ especially in the presence of a dispersive f-band ($t_f>0$). As our primary goal is to study corrections due to hybridization on a dispersive model we have set $\Delta_T=t_c$ for the localized band and $\Delta_T=t_f$ for the dispersive systems studied below.

Our search for superconducting order due solely to hybridization began in the PAM models where we evaluated susceptibilities of the model for the cases with local ($V_0$), nearest neighbor ($V_1$) and frustrated hybridization ($V_0+V_1$), as the system is doped away from half filling. The response of the system for the three cases is shown in Fig.\ref{frustrated_SU}. Antiferromagnetic fluctuations are dominant at half filling for the models without frustration. Doping leads to the destruction of the nested Fermi surface, and we find the AF order is suppressed at finite doping for all cases. As we work primarily at moderate coupling ($U\sim 0.25W_c$) the energy of interest is the exchange coupling $J\sim V_0^2/U$. For small $V_0$ the exchange coupling stabilizes the AF order but as the local hybridization is increased the conducting band begins to screen the moments leading to the suppression of the spin order. This screening of the local moments in the 2D PAM is dependent on the local Hubbard coupling, hybridization and temperature of the system. Fig.\ref{2DPAM_v0T} shows the temperature dependence of the ferromagnetic response of the 2D PAM model for different levels of hybridization. At high temperatures the spin response goes as $\chi_{FM}\sim 1/T$ for all values of $V_0$ but as we decrease the temperature we find a faster screening of the moments that scales with hybridization with the conduction band. Suppression of the AF and FM responses with increasing $V_0$ can be seen in the top panel of Fig.\ref{frustrated_SU}.  Moving the chemical potential of the localized band can lead to a reemergence of the local moments, although the FM response is still sensitive to $V_0$ and the value of the local repulsion. Unlike local hybridization which screens spin response, the nearest neighbor hybridization ($V_1$) stabilizes the AF over a large range of dopings and presents a spin response similar to that observed in the Hubbard model. The response of the PAM to nearest neighbor hybridization ($V_1$) is shown in middle panel of Fig.\ref{frustrated_SU}. Large doping leads to a suppression of both the AF and FM responses with s-type superconductivity emerging as the leading order at $V_1\sim t$ and large doping ($p>0.2$). The presence of both local and nearest neighbor hybridization can frustrate the system ($V_1\rightarrow V_0$) leading to the suppression of antiferromagnetic fluctuations and emergence of ferromagnetism at half filling. The ferromagnetic order is unstable to doping and quickly gives way to a metallic phase. Doping allows for antiferromagnetic correlations to reemerge from the frustrated system with $s-type$ superconducting order occupying the region between the antiferromagnetic and ferromagnetic states.

\begin{figure}
    \centering
    \hspace*{-0.25cm}\includegraphics[scale=0.4]{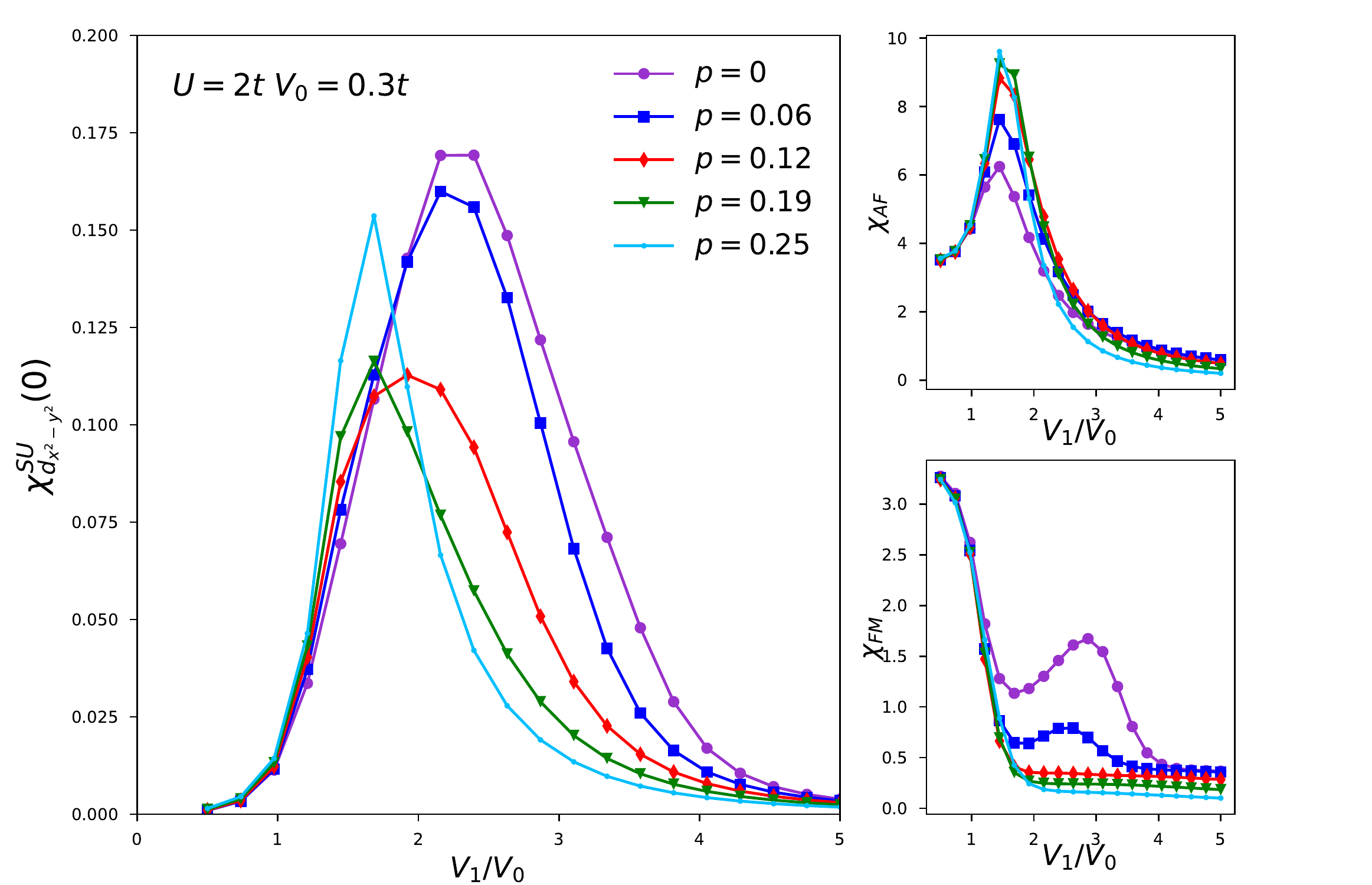}
    \caption{The $d_{x^2-y^2}$ superconducting (left), antiferromagnetic (right) and ferromagnetic response of the 2D PAM, $U=2t$ $W_c=8t$, with local ($V_0=0.3t$) and nearest neighbor hybridization as a function of the hybridization ratio ($V_1/V_0$) for different values of doping.}
    \label{2DPAM_vRatio}
\end{figure}

\begin{figure*}
  \centering
  \hspace*{-0.2cm}\includegraphics[scale=0.4]{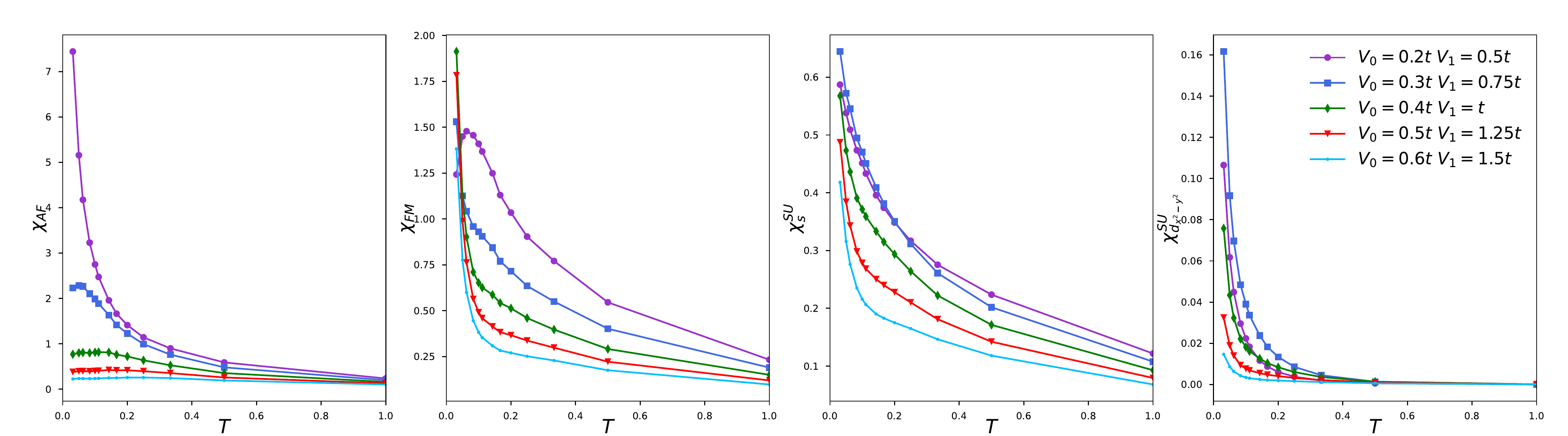}
  \caption{The antiferromagnetic, ferromagnetic, superconducting correlations ($s$, $d_{x^2-y^2}$) of the 2D PAM, $U=2t$ $W_c=8t$, as a function of temperature for different values of hybridization ($V_0$) at a fixed ratio local and nearest neighbor hybridization ($V_1/V_0=2.5$).}
\label{tremblay_SU}
\end{figure*}

Frustration can enhance superconducting correlations with previous studies finding both the ratio ($V_0/V_1$) and hybridization ($V_0$) to play a role in the enhancement\cite{wu2015d}. The interplay between local and nearest neighbor hybridization and its impact of the superconducting response of the 2D PAM model is shown in Fig.\ref{2DPAM_vRatio}. The $d_{x^2-y^2}-SU$ response is suppressed for local ($V_0$) and nearest neighbor hybridization ($V_1$) but shows enhancement for $V_1/V_0\sim 2$ for a variety of doping values. For a half filled system the response is peaked at $V_1/V_0\sim 2.5$ which was used to set the ratio for Fig.\ref{tremblay_SU}, which explores the dependence on temperature and hybridization ($V_0,V_1$) of the frustrated system. The AF response is strongest for $V_0=0.2t$ and uniformly decreases with increasing $V_0$. Alternatively, the FM response which is screened for small $V_0$ is enhanced with increasing hybridization with the system showing dominant FM correlations for $V_0>0.3t$. The point of transition from AF to FM order shows the strongest superconducting response with both $s$ and $d_{x^2-y^2}-SU$ response peaking at $V_0\sim 0.3t$. The results are in accordance with previous works and highlight the emergence of superconductivity with frustration and doping\cite{wu2015d}.

\section{The Three-dimensional Periodic Anderson Model}\label{3DPAM}

Although a two dimensional PAM can be used to model the role of hybridization in the Nickelates, recent proposals have suggested the hybridization with the interstital-s band plays an additional role as the driver of a crossover from two to three dimensional superconductivity. Investigating the mechanism and the crossover requires treating the three-dimensional structure of the hybridization. Hence, in this section, we investigate, along with hybridizations to local and nearest neighboring sites, the impact of hybridization with the interstitial-s band present in the Nickelate compounds. Following our approach to the 2D PAM, we examine the impact of doping on the ferromagnetic, antiferromagnetic and superconducting correlations present within these systems. The resolutions for the fRG used to construct the flow were set at $N_k=1$, which corresponds to 7 basis functions, and $N_\omega=4$ with $\Delta T=t$. To account for the larger bandwidth of the 3D conducting band the local Hubbard coupling has been adjusted to keep $U/W_c=0.25$ fixed.

\begin{figure}
    \centering
    \includegraphics[scale = 0.295]{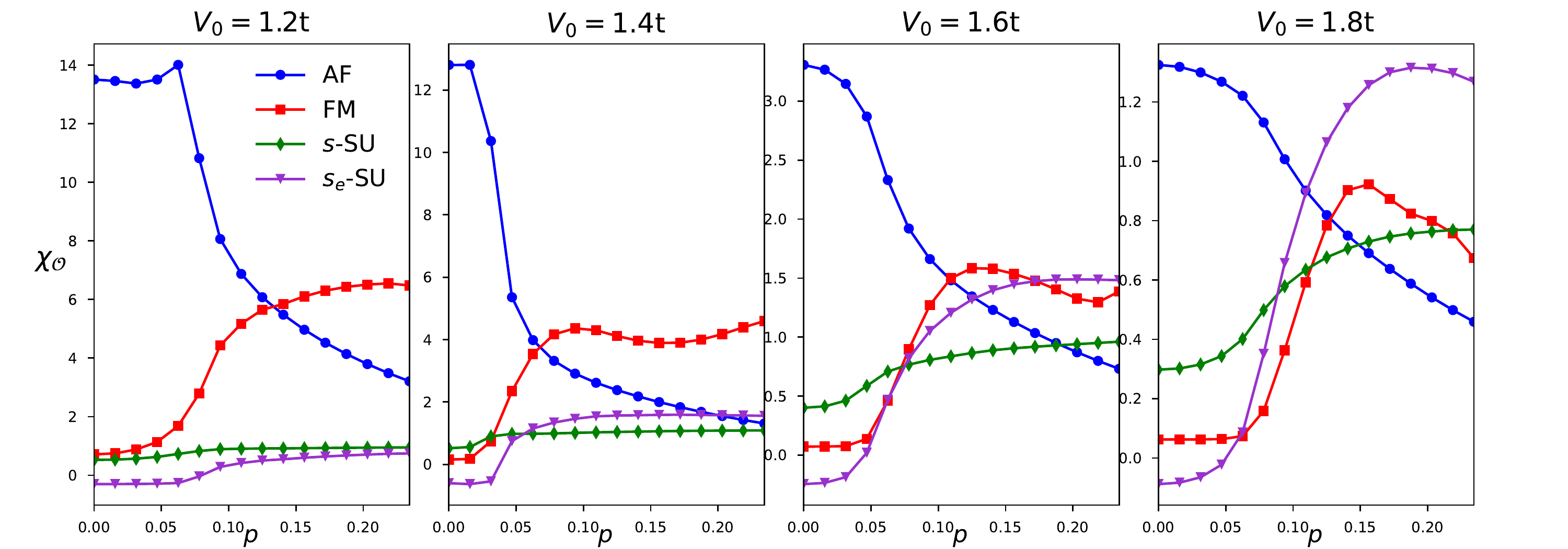}
    \includegraphics[scale=0.295]{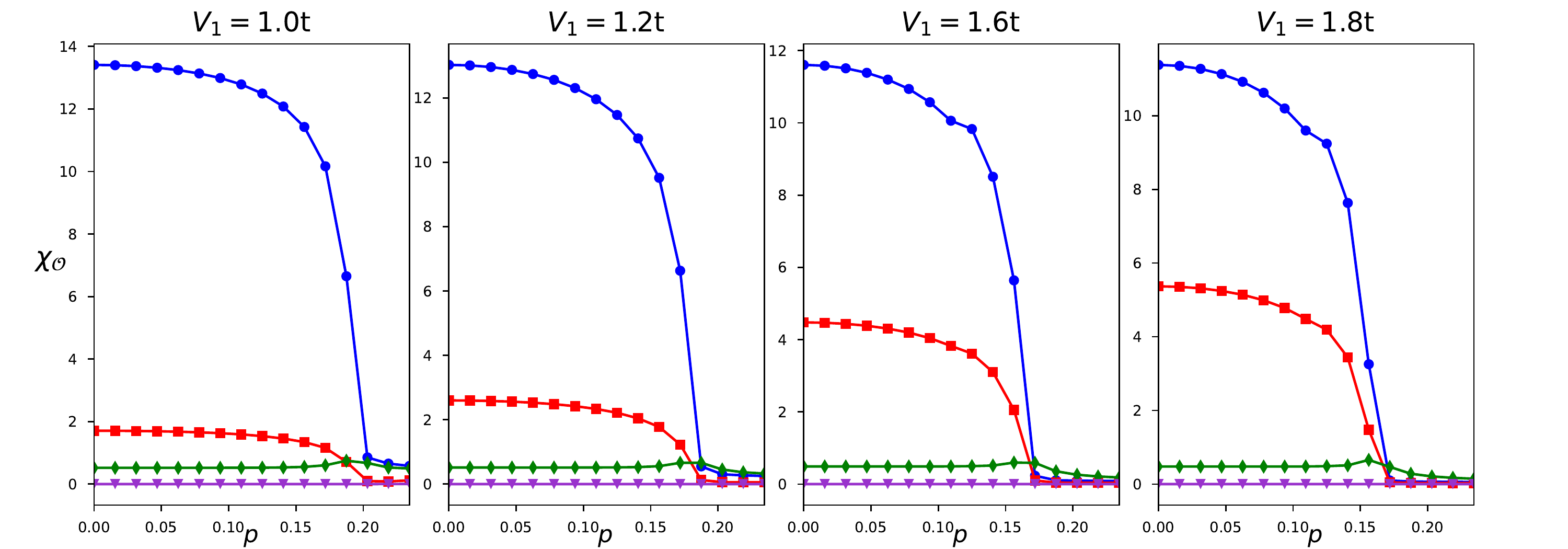}
    \includegraphics[scale=0.295]{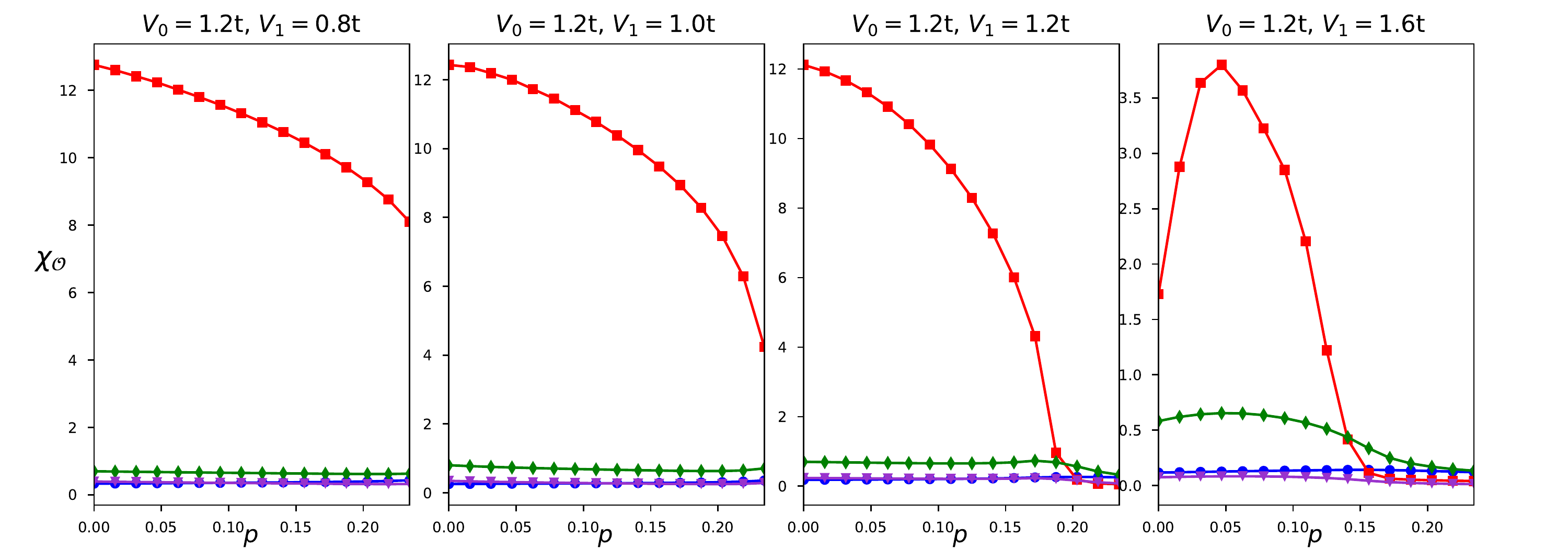}
    \caption{Antiferromagnetic (AF), ferromagnetic (FM), and superconducting ($s-SU$, $s_e-SU$) correlations of the 3D PAM, $U=3t$ $W_c=12t$, on an $8\times 8\times 8$ lattice with local hybridization ($V_0$), nearest neighbor hybridization ($V_1$) and nearest neighbor hybridization ($V_1$) in the background of local hybridization ($V_0=1.2t$) as a function of doping are shown in the top, middle and bottom panels. We used a 2-loop fRG with $N_\omega=4$, $N_k=1$ and $\beta=32$.}
    \label{3Dspin}
\end{figure}

\begin{figure}
\centering
\includegraphics[scale=0.58]{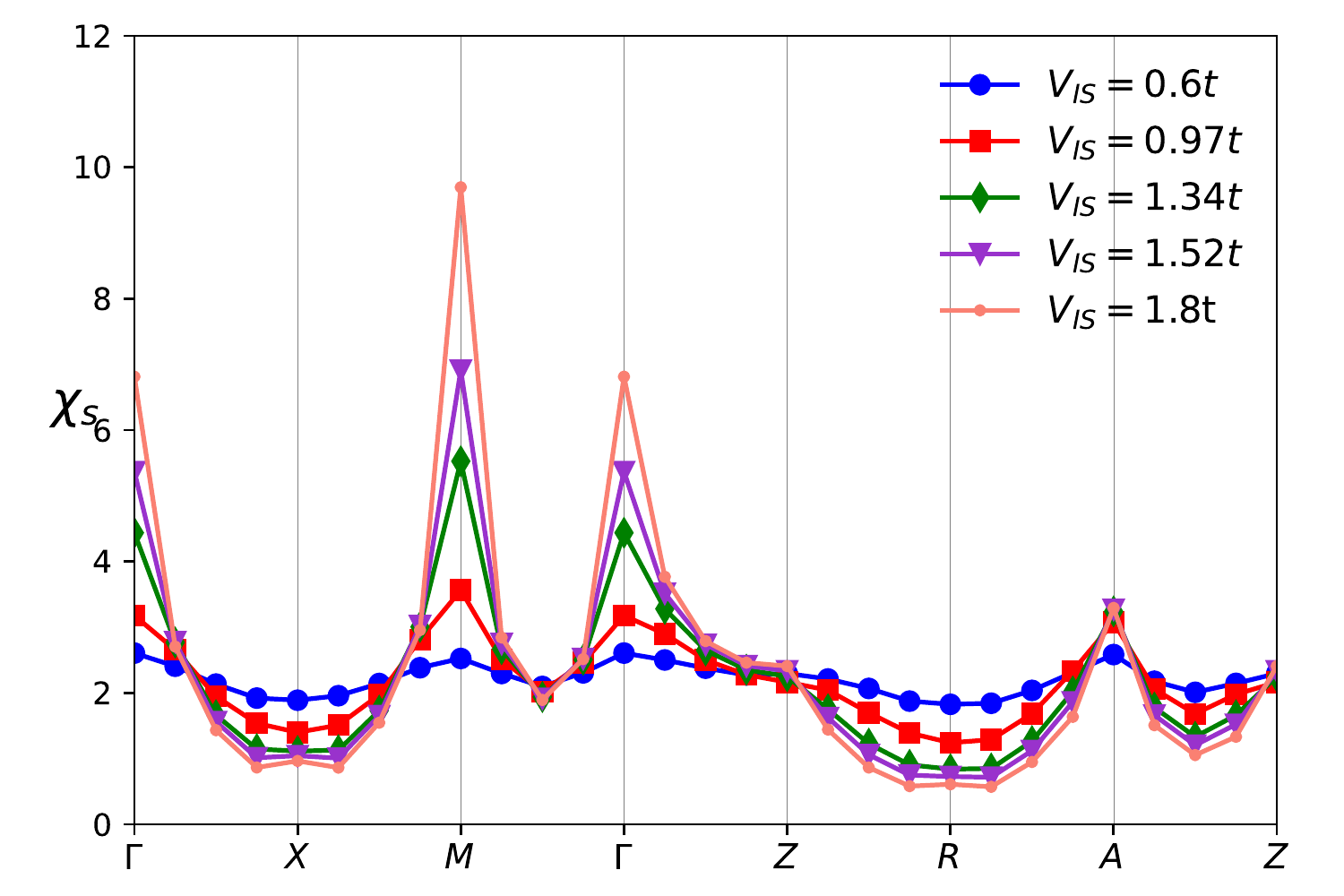}
\caption{The spin response of the 3D PAM hybridized with an interstitial-$s$ band ($V_{IS})$ with $U=3t$, bandwidth $W_c=12t$ on an $8\times 8\times 8$ lattice at $\beta=16$.}
\label{af3DPAMVS}
\end{figure}

The response of the doped 3D PAM to different values local and nearest neighbor hybridizations is shown in Fig.\ref{3Dspin}. As in the 2D PAM case, the interplay between the Hubbard coupling and local hybridization can be captured by the exchange coupling given by $J\sim V_0^2/U$ which for large values of $V_0$ leads to a suppression of the spin response at all values of doping. The differences to the 2D PAM response revolve around the reduced the spin response and the faster suppression with doping both of which can be explained due to the changes in the dimensionality of the system. Beyond the local ($s$-wave) superconducting response $V_0$ also drives an extended s-wave which unlike the $s$-wave response is enhanced with increasing $V_0$. For low values of $V_0$, the AF order is stable but as the hybridization is increased we can expect the system to transition to a metallic phase with strong superconducting correlations. Unlike the local case the nearest neighbor hybridization appears to replicate the physics of the 3D Hubbard model with $V_1$ simply controlling the strength of AF response and we observe the expected suppression of the spin response as the system is doped. The last row of Fig. \ref{3Dspin} shows that the effect of frustration in the lattice which enhances the ferromagnetic response, similar to the behavior in the 2D model. The FM response weakens with increasing $V_1$ with the system finally transitioning to the expected metallic phase.

\begin{figure}
\centering
\includegraphics[scale=0.44]{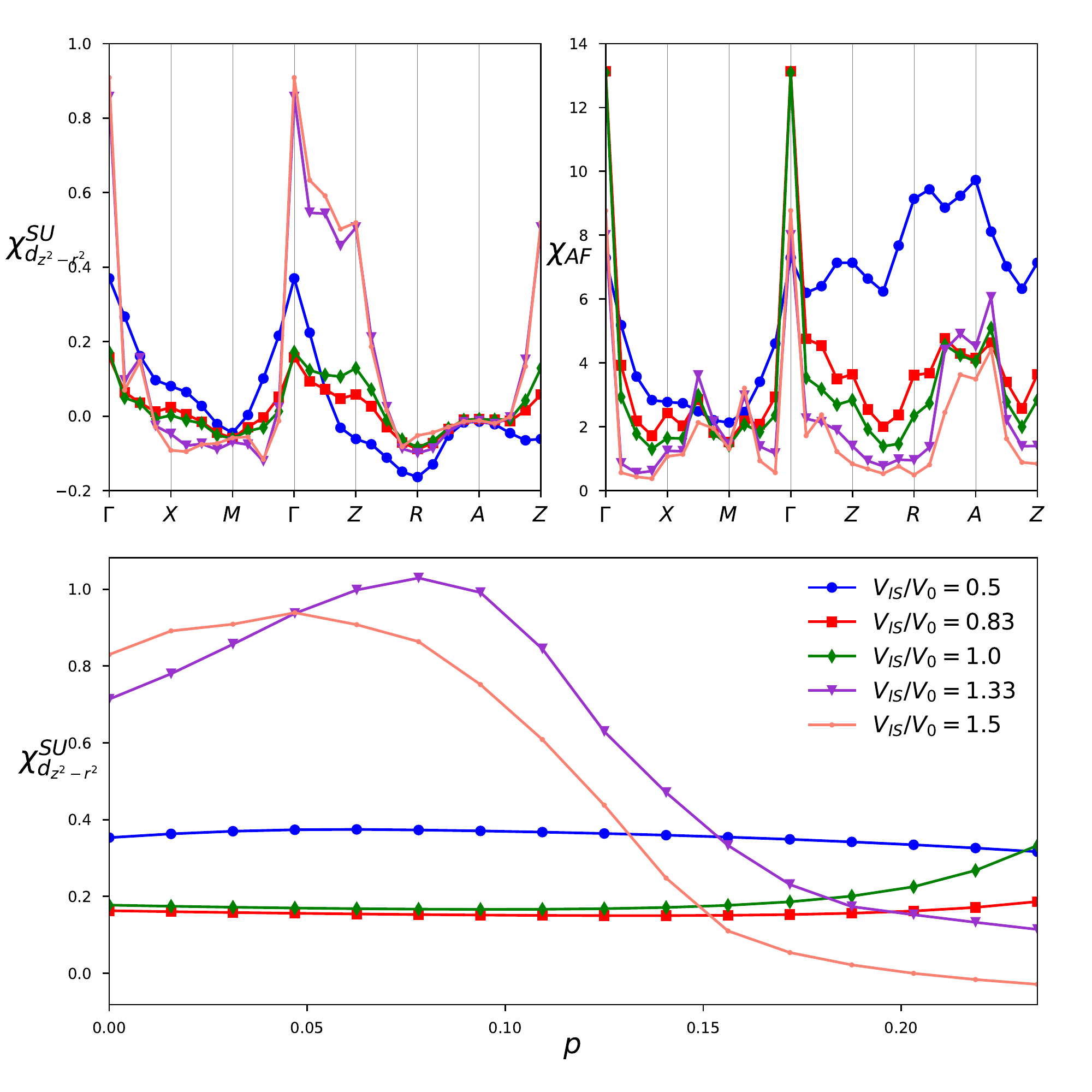}
\caption{Superconducting and spin response of the 3D PAM hybridized locally ($V_0$) and with an interstitial-s band ($V_{IS}$). The momentum dependence of the $d_{z^2}$-superconducting and spin susceptibilities for different hybridizations ($V_{IS}/V_0$) is shown in the top panels. The doping dependence of the $d_{z^2}$-superconudcting response is shown in the bottom panel.}
\label{momDep3DV0VISres}
\end{figure}

Unlike locally hybridized systems, hybridization with the interstitial s band, defined in Eq.\ref{hybridizationVIS}, does not lead to the complete screening of the spin response in the system. The response observed is similar to the case for nearest neighbor hybridization with the system showing enhanced in-plane ($(\pi,\pi,0)$) and ferromagnetic spin response with increasing hybridization. The spin response of the 3D PAM hybridized with the interstitial $s$-band is shown in Fig.\ref{af3DPAMVS}. We can introduce frustration in the system by allowing for local hybridization ($V_0\neq 0$), which biases the system towards a 3D AF. The spin and superconducting responses of the system for different degrees of frustration are shown in Fig.\ref{momDep3DV0VISres}. Much like the frustrated 2D PAM, we see enhanced superconducting correlations between the AF and ferromagnetic spin regimes. The observed superconducting response has a $d_{z^2-r^2}$ symmetry and peaks at finite doping. Though the lack of dispersion in the localized band presents an incomplete picture, hybridization with the interstitial-$s$ does suppress the 3D antiferromagnetic response ($\pi,\pi,\pi$) and in the presence of local hybridization shows enhanced 3D superconducting fluctuations.

\section{Incorporating dispersion and hybridization in the Nickelates}
\label{NickelSec}

The observation of superconductivity in the Nickelates represents the culmination of a long search for Cuprate analogs in the quest to understand high-$T_c$ superconductivity\cite{anisimov1999electronic}. Superconductivity was observed in $\mathrm{Sr_{x}Nd_{1-x}NiO_{2}}$ and $\mathrm{Sr_{x}Pr_{1-x}NiO_{2}}$ films in the doping range, $x$ from 0.12 to 0.28 with a transition temperature, $T_c$, in the $9-15K$ range\cite{li2019superconductivity,pickett2021dawn,chow2022pairing,osada2020phase}. The undoped parent compounds are of the type $\mathrm{RNiO_{2}}$, where the specific rare earth metals R=Nd,Pr have shown superconductivity, both having infinite layers of square $\mathrm{NiO_{2}}$ planes with $3d^9$ electron configuration for the $\mathrm{Ni}$. Though similar to the Cuprates with a $d_{x^2-y^2}$ band crossing to match, the analogy is complicated by lack of antiferromagnetic order in $\mathrm{NdNiO_{2}}$ and the presence of a $d_z$ orbital due to the Nd-bands. The pocket created by the Nd-5d orbitals leads to self-doping effects as electrons are removed from the $\mathrm{Ni}-d_{x^2-y^2}$ to the Nd-5d pockets even for the undoped  $\mathrm{NdNiO_{2}}$ which may explain the lack of insulating AF normally seen in the Cuprates\cite{nomura2022superconductivity}. The charge transfer energy of the Nickelates is also higher than the Cuprates, which suggests dominant Mott-Hubbard physics. Hence, modeling efforts have focused on the d-bands at the Fermi surface to capture the competing orders present in the system\cite{karp2020comparative}. Long range AF order is absent in the doping range of interest ($x=0.12-0.28$) with the superconducting order emerging from a possible charge ordered state\cite{tam2022charge,rossi2022broken,shen2023electronic}. The charge ordering observed in the $NdNiO_2$ system is intricate and appears to form in the $Nd$ and $Ni$ orbitals at the same nesting vector, with other studies finding less exotic charge ordering along the $Ni$ bonding direction. The impact of the self-doping orbital on the charge and superconducting response present in the system remains unexplored, although the difference in $T_c$, $~10K$ for the Nickelates and $~110K$ for the Cuprate analog($\mathrm{CaCuO_2}$), suggests an overall negative impact\cite{botana2020similarities}.

\begin{figure}
\centering
\includegraphics[scale=0.55]{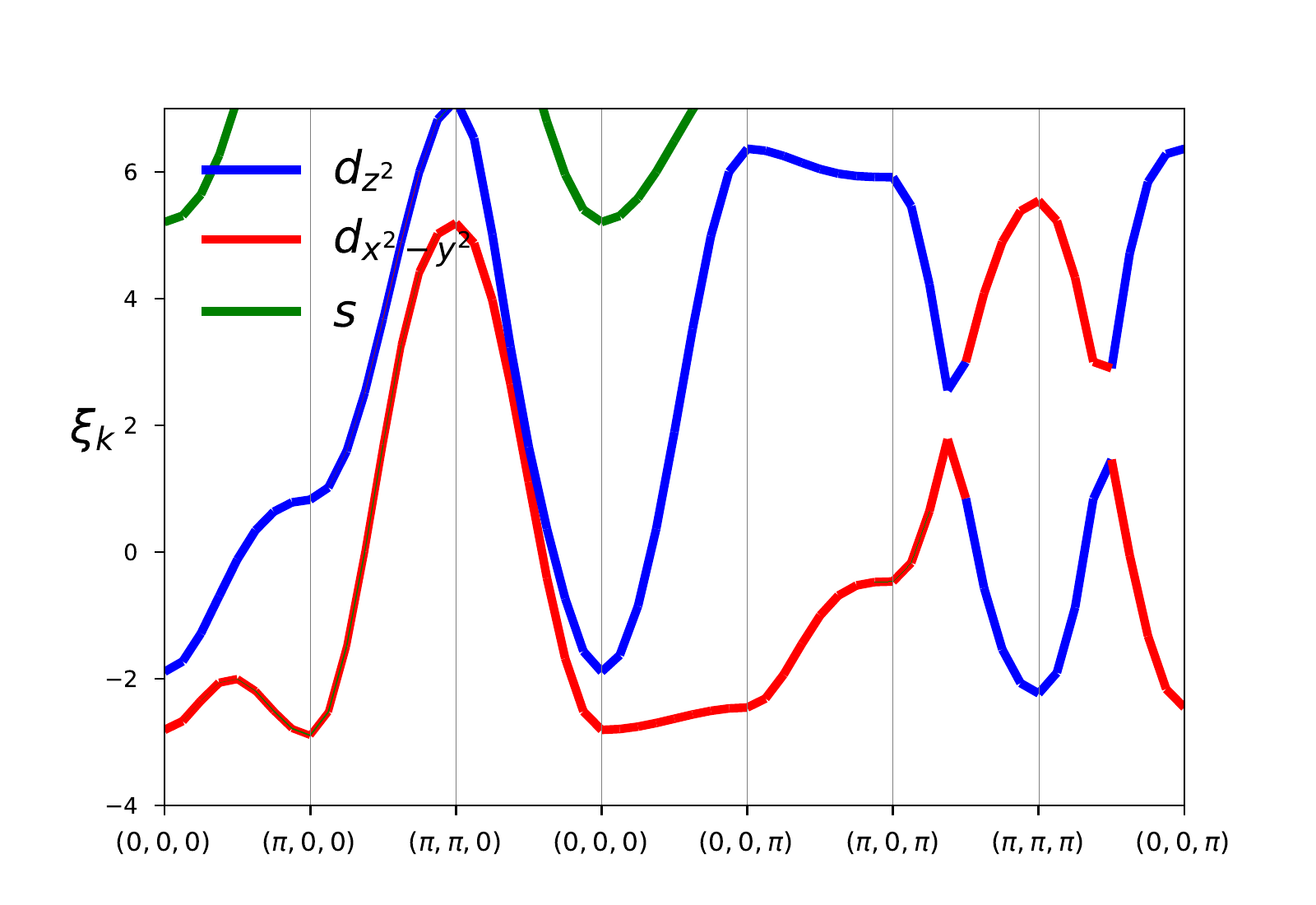}
\caption{Band Structure for an {\it ab-initio} two-orbital model of the Nickelates\cite{been2021electronic}. The hopping was normalized by the Ni $d_{x^2-y^2}$-hopping in the x/y directions. The interstitial s orbital hybridizes only with the Ni-d orbital.}
\label{bandStrucNic}
\end{figure}

\begin{figure}
\centering
\includegraphics[scale=0.43]{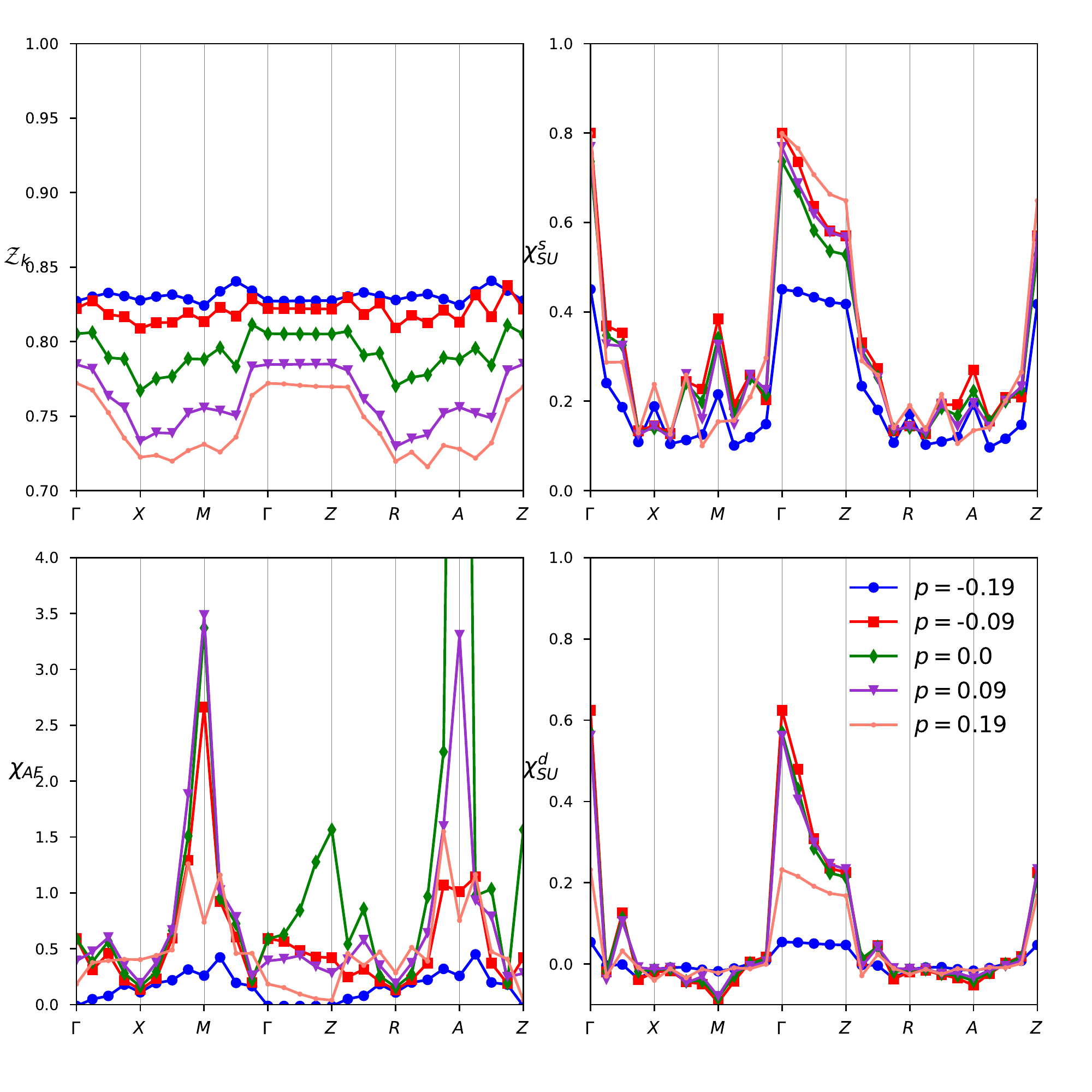}
\caption{Momentum dependence of the quasi-particle weight, the spin response and the superconducting susceptibilities of a model for $NdNiO_2$ at different values of doping for $U=0.3W_{Nd}$.}
\label{afNdNickUdepBZ}
\end{figure}

\begin{figure*}
\centering
\includegraphics[scale=0.4]{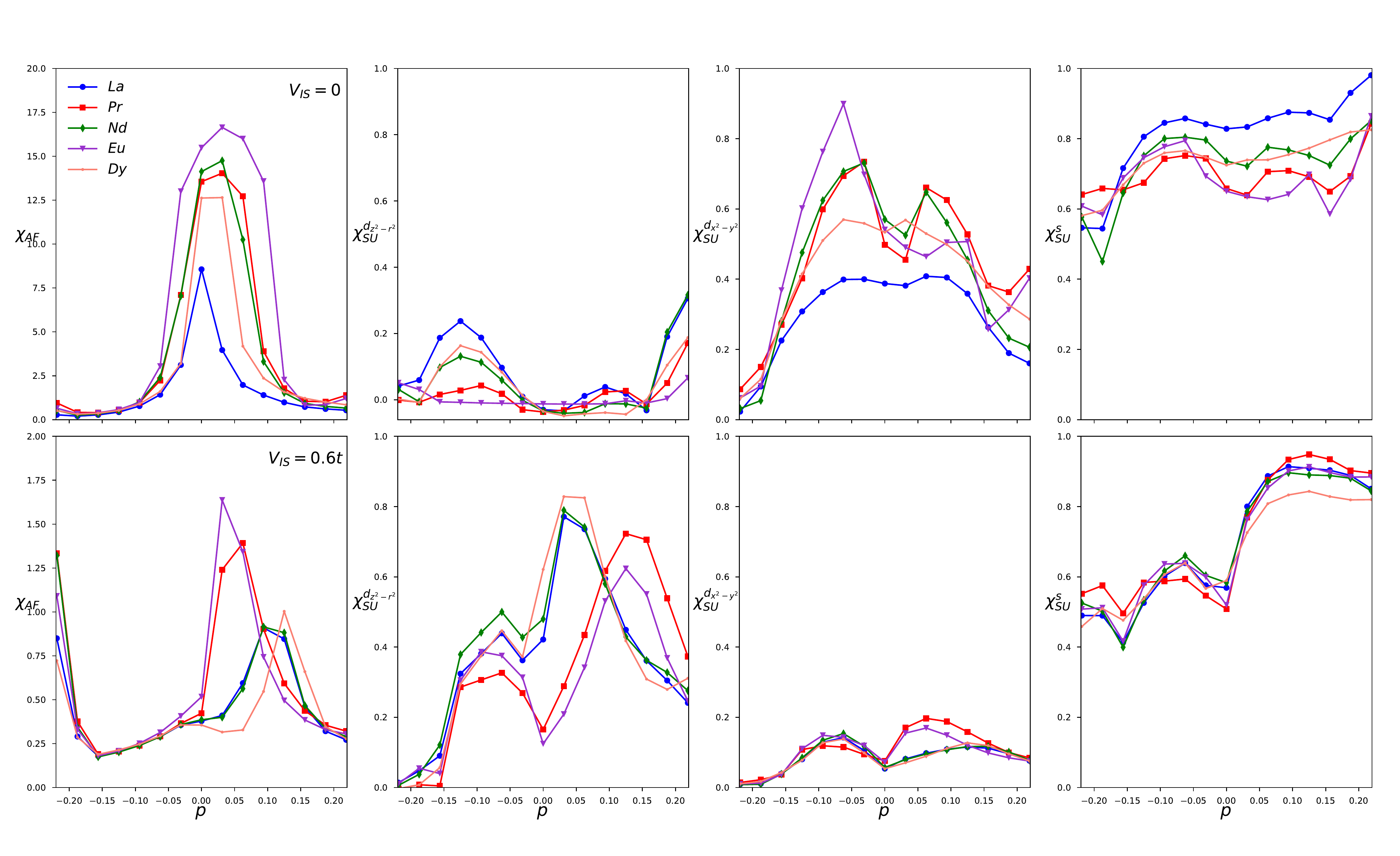}
\caption{The antiferromagnetic, $d_{z^2-r^2}$, $d_{x^2-y^2}$, and $s$ superconducting susceptibilities as a function of doping for the ab-initio two orbital model of the Nickelates with various rare earth elements hybridized with the interstitial-s orbital ($V_{IS}=0,0.6t$). Calculations were performed at 2-Loop on an $8\times 8\times 8$ lattice for $\beta=32$, $N_\omega=4$ and $N_k=1$ with a Hubbard interaction $U=0.3W_{Ni}^R$.}
\label{susNickSeries}
\end{figure*}

To understand the origin of superconductivity in the Nickelates, their electronic structure has been thoroughly studied\cite{ryee2020induced,lechermann2020late}. The consensus as to the key ingredient driving the ordering is the $\mathrm{Ni}-d_{x^2-y^2}$ orbital, which contributes most of the weight at the Fermi level. Further, the large charge transfer energy of the Nickelates and the small contribution from the $\mathrm{Ni}-d_{z^2}$ at the Fermi level suggests a minimal impact of the $\mathrm{O}-p$ orbitals and the $\mathrm{Ni}-d_{z^2}$ orbitals on the overall picture\cite{hepting2020electronic}. The other band at the Fermi surface is the $\mathrm{Nd}-5d_{z^2}$ which forms a pocket around the $\Gamma$ point and hybridizes weakly with $\mathrm{Ni}-d_{x^2-y^2}$. Although the contribution of this orbital is included, there is debate as to the strength of hybridization ($V$) between the bands. A lack of electronic correlations and a weak hybridization would relegate the impact of the band to the ordering in the system to the single particle level with its function restricted to that of a charge reservoir. In contrast, a moderate hybridization would introduce screening due to Kondo physics and might explain the lack of an insulating state in the Nickelates. Differing levels of occupation for the different Nickelate compounds across the the rare earth atoms R can be crucial in identifying the role of the $5d_{z^2}$ band in the Nickelates\cite{been2021electronic}. The final piece to consider is the interstitial-s orbital, which previous works have argued hybridizes strongly with the $\mathrm{Ni}-d_{x^2-y^2}$ orbital and works to screen the local $Ni$ moments\cite{gu2020substantial}. In the sections below, we explore the role of hybridization with these bands and its impact on the system's response at various doping levels by constructing the fRG flow for {\it ab-initio} and model Hamiltonians of the Nickelates.

\subsection{{\it Ab-initio} models}

\begin{figure*}
\centering
\includegraphics[scale=0.39]{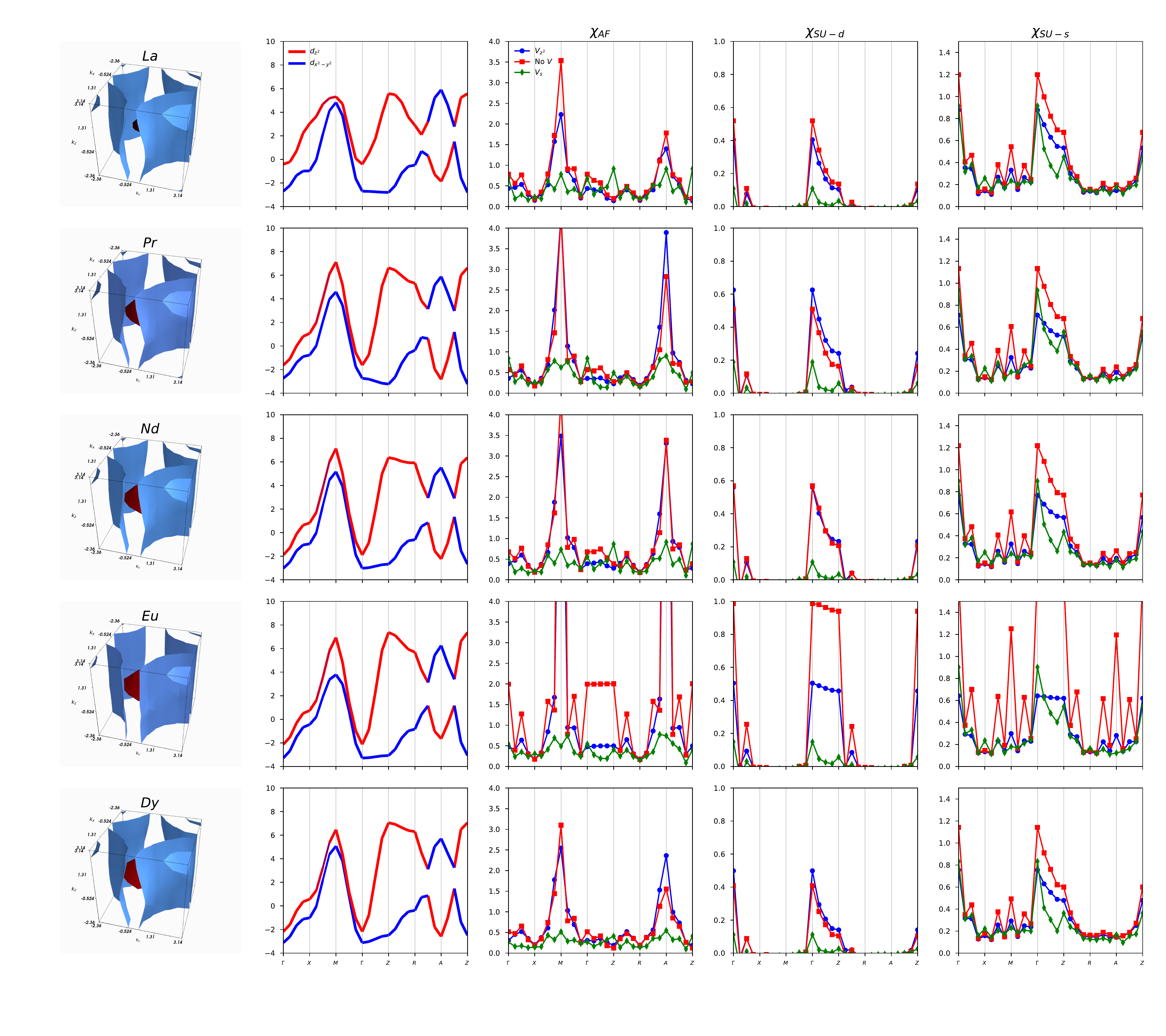}
\caption{The Fermi surface, band structure, spin and superconducting susceptibilities of the hole doped ($p=0.09$) rare earth series of Nickelates. The response functions for models with the Ni-$d_{x^2-y^2}$ orbital without hybridization as well as with hybridization with the Nd-$d_{z^2}$ and interstitial-$s$ orbital are shown.}
\label{nickSeriesFSsus}
\end{figure*}

In this section, we consider {\it ab-initio} models constructed to capture a family of RNiO$_2$ systems\cite{been2021electronic,hepting2020electronic}. Tuning the rare earth R in this family is of considerable current interest, as the reported $T_c$ of $2K$ for R=La is an order of magnitude smaller than the $T_c=10-15K$ observed in the R=Nd,Pr compounds, and the ionic size of the rare earth layer appears to control the dimensionality of the superconducting order\cite{zeng2022superconductivity,chow2023dimensionality}. Studies analyzing the role of the R layer can not only help offer prescriptions for enhancing $T_c$ in the system, but due to variations in the hybridization, charge transfer energies, and occupation of the R-$5d$ orbital, they can also help in identifying the model that most readily captures superconductivity in the system.  The model Hamiltonian for which the {\it ab-initio} parameters have been determined consists of the correlated $3d_{x^2-y^2}$-Nickelate orbital hybridizing with a conducting band having mostly $\mathrm{Nd}-5d_{z^2}$-character. The synthesis of $RNiO_2$ requires the removal of the apical oxygen in $RNiO_3$ leading to the formation of an interstitial-$s$ which is modeled here as independent of R, but generally we can expect the value of the hybridization ($V_{IS}$) and the relative shift ($\xi=8t_R$) to ultimately depend on the rare earth metal in the compound. The parameters of the single particle Hamiltonian associated with the Ni-$d_{x^2-y^2}$ and R-$d_{z^2}$ are as given in Ref.\onlinecite{been2021electronic}. The parameters include contributions from the low lying Ni-$d_{z^2}$ and R-$d_{xy}$ orbitals. The hybridization between the two bands is of the form
\begin{align}
V_k=-8V_d\cos(k_z/2)\bigl(\sin(3k_x/2)\sin(k_y/2)-\nonumber\\
\sin(k_x/2)\sin(3k_y/2)\bigl).
\end{align}
with $V_d$ set to the value given in Ref.\onlinecite{been2021electronic}. The hybridization is relatively weak, going roughly as $V_d=0.05-0.1 t_{Ni}^R$. The hybridization of the Ni-$3d_{x^2-y^2}$ band with the interstitial-s if given above in Eq.\ref{hybridizationVIS} and involves the narrower s-band ($W_{Ni}\sim 2W_{IS}$) left empty with $\xi_{IS}=8t_R+1.2\sum_i^3\cos(k_i)$. The hybridization to the interstitial-$s$ is substantial and reported to be an order of magnitude stronger than $V_d$ ($V_{IS}=0.5t_{Ni}^R$)\cite{gu2020substantial}. The dispersing PAM Hamiltonian that corresponds to the {\it ab-initio} values is solved via the decoupled fRG for at moderate coupling
($U=0.5W_{Ni}^{R}$). To counter the limitation of the fRG to moderate coupling, we study the response of the system at various Hubbard couplings and extrapolate the system's response to the Nickelate parameter regime. For the values considered, the dependence on the interaction appears to be related only to the magnitude of the response, and as the parent compound of the Nickelates is expected to be a bad metal, we expect the moderate coupling regime to be adiabatically connected to the Nickelate system. In what follows, we study the effects of hybridization between Ni-d, R-d and interstitial-$s$ bands, the sensitivity of the ordering tendencies to the degree of correlation ($U$), and identify doping ranges that show novel ordering in the system.

The band structure for a minimal model of the superconducting $\mathrm{NdNiO_2}$ is shown in Fig.\ref{bandStrucNic}. Notable features are the weak out of plane dispersion appearing as a plateau from $(0,0,0)$ to $(0,0,\pi)$ as well as the pocket formed by the Nd-$d_{z^2}$ orbital at $(0,0,0)$. The interstitial-$s$ orbital lies higher in energy ($\Delta_E=3 eV$), and hybridizes strongly with the Ni-$d_{x^2-y^2}$ orbital ($V_s=0.22 eV$, $V_{z^2}=0.02eV$) which contributes to the extension of the system beyond 2D $NiO_2$ planes. Working in the moderate coupling regime ($U\sim 0.3W$), we looked at the effect of doping on the response in the system. The doping range of interest for the thin film superconducting samples of Sr doped PrNiO$_2$ ($T_c=13K$) is $p=0.1-0.3$ \cite{nomura2022superconductivity}. Given the weak degree of hybridization, we can expect the pocket of the rare earth (R) in the Nickelates to provide different levels of screening across the doping range for the series of compounds.
\begin{figure}
\centering
\includegraphics[scale=0.4]{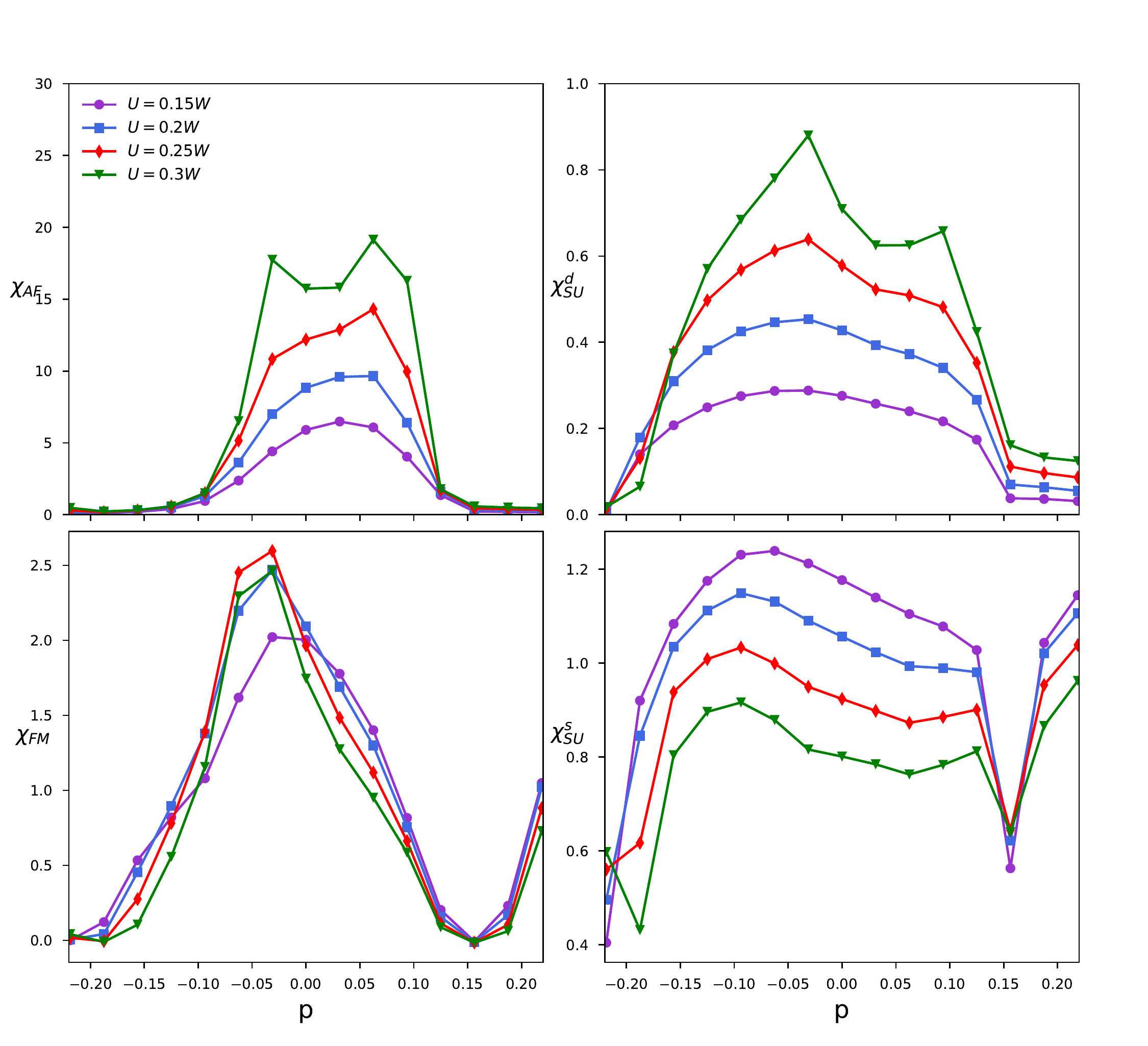}
\caption{The spin and superconducting susceptibility of the 2D Nickelate model as a function of doping for different strength of the Hubbard interaction.}
\label{afNick2DUdep}
\end{figure}

Despite the lack of long range AF order in the Nickelates in the doping range associated with superconductivity, we observe a relatively strong AF response suggesting spin fluctuations as the primary drivers of superconductivity in the system. The doping dependence of the spin and superconducting susceptibilities for the $Nd$-Nickelates is shown in Fig.\ref{afNdNickUdepBZ}. We see that the AF response is suppressed at large electron doping, with the peak emerging at M ($\pi,\pi,0$) and A ($\pi, \pi, \pi$) indicating a planar AF response as we approach the hole doped regime. Further, doping ($p>0.15$) leads to an incommensurate spin fluctuations which weakens the $d_{x^2-y^2}$-superconducting response indicating a possible endpoint to the doping range of interest. Overall the s-SU response shows little change cross the relevant doping range and persists even as incommensurate spin fluctuations emerge at high hole doping. The presence of finite out of plane hopping ($t_z\neq 0$) combined with the coupling of strong planar spin fluctuations in the Ni-band to the metallic-R pocket leads to a reduced planar response and raises questions as to nesting vector associated with the AF response. Across the series, the Nickelates show strong spin fluctuations at
M ($(\pi,\pi,\pi)$) and at A ($(\pi,\pi,0)$) with the series showing a preference for a $(\pi,\pi,\pi)$ response for the La,Dy-Nickelates and switching back to a quasi-2D response at M ($\pi,\pi,0$) and A ($\pi, \pi, \pi$) for the remaining Nickelates. Looking at the quasi-particle renormalization of the Nd system shown in Fig.\ref{afNdNickUdepBZ} we see reductions around $X(\pi,0,0)$ and $R(\pi,0,\pi)$ indicating a bias towards a planar configuration. The difference in the nesting vector can possibly be attributed to the reduction of the out-of-plane lattice constant ($t_{Ni}^{[0,0,1]}$) which dips across the series \cite{been2021electronic}. Changes in the relative strength of the AF response combined with differences in the sensitivity of spin and superconducting fluctuations to doping across the series suggest the rare earth bands have a strong indirect effect beyond a simple screening of the spin fluctuations.

The trends in the AF and superconducting responses seen across the series is shown in Fig.\ref{nickSeriesFSsus}. The calculated system response is strongest for the Pr, Nd and Eu systems   across the board with the Eu compound showing a 2D response in the absence of hybridization with the interstitial-$s$. Of particular note is the variation seen in the response for the same Hubbard interaction strength ($U/W=0.25$) despite similar parameters ($t_z\sim 0.1$) for the Ni band. This appears primarily due to the contributions of the rare earth band, which can be further explored by varying the filling independently of the Ni band. Preliminary work along this direction is shown in Sec.\ref{gPAM} where the Ni band is modeled by a fixed 2D Hubbard model as the hybridization with the rare earth band is varied. The strength of the superconducting fluctuations is closely tied to the electronic coupling, with larger $U$ leading to a stronger AF response which in the doped regimes serve to mediate superconductivity. Less clear are the parameters driving the response in the Eu system. Parameters across the models such as the hybridization and Ni levels trend similarly, which suggests that the increasing Nd energy levels lead to a decoupling at larger values ($\xi_{Eu}=1.36ev$, $\xi_{Dy}=1.41ev$). Though hybridization with the rare earth band suppresses spin response at $\Gamma$ it does not affect the long range antiferromagnetic response which is absent in the Nickelates. Hybridization with the interstitial-S appears is crucial for the overall suppression of the spin response though this is accompanied by a significant reduction in the $d-SU$ from the unhybridized response shown in Fig.\ref{nickSeriesFSsus}. The figure catalogues the impact of $V_d$ and $V_{IS}$ on the antiferromagnetic and superconducting response of the rare earth Nickelates at $p=0.09$. The suppression of d-type superconductivity and AF response in the $V_{IS}$ hybridized system is accompanied by an s-type superconducting response on par with the unhybridized system. The spin, charge and superconducting responses across the entire doping range for the Nickelate model hybridized with the interstitial-s are shown in Fig.\ref{susNickSeries}. We see that Eu and Pr show a strong predicted $s,d_{z^2}$ superconducting response, which should be of interest to experimental studies.

\begin{figure}
\centering
\includegraphics[scale=0.6]{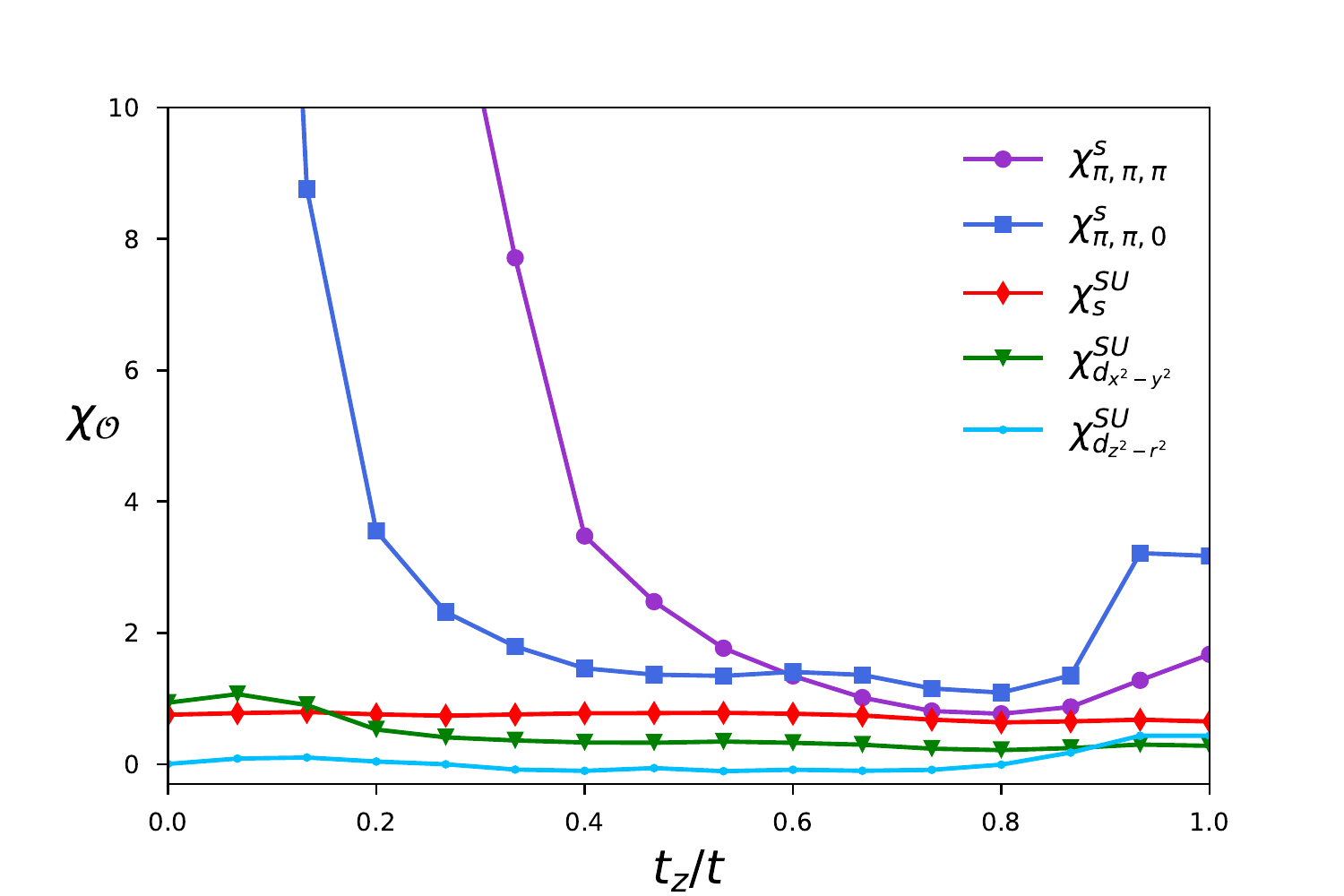}
\caption{The spin and superconducting susceptibilities of the quasi-2D Hubbard model ($U=0.3W$) as a function of the out of plane hoping at fixed doping ($p=0$).}
\label{susNick3DzUdep}
\end{figure}

\begin{figure*}
\centering
\includegraphics[scale=0.35]{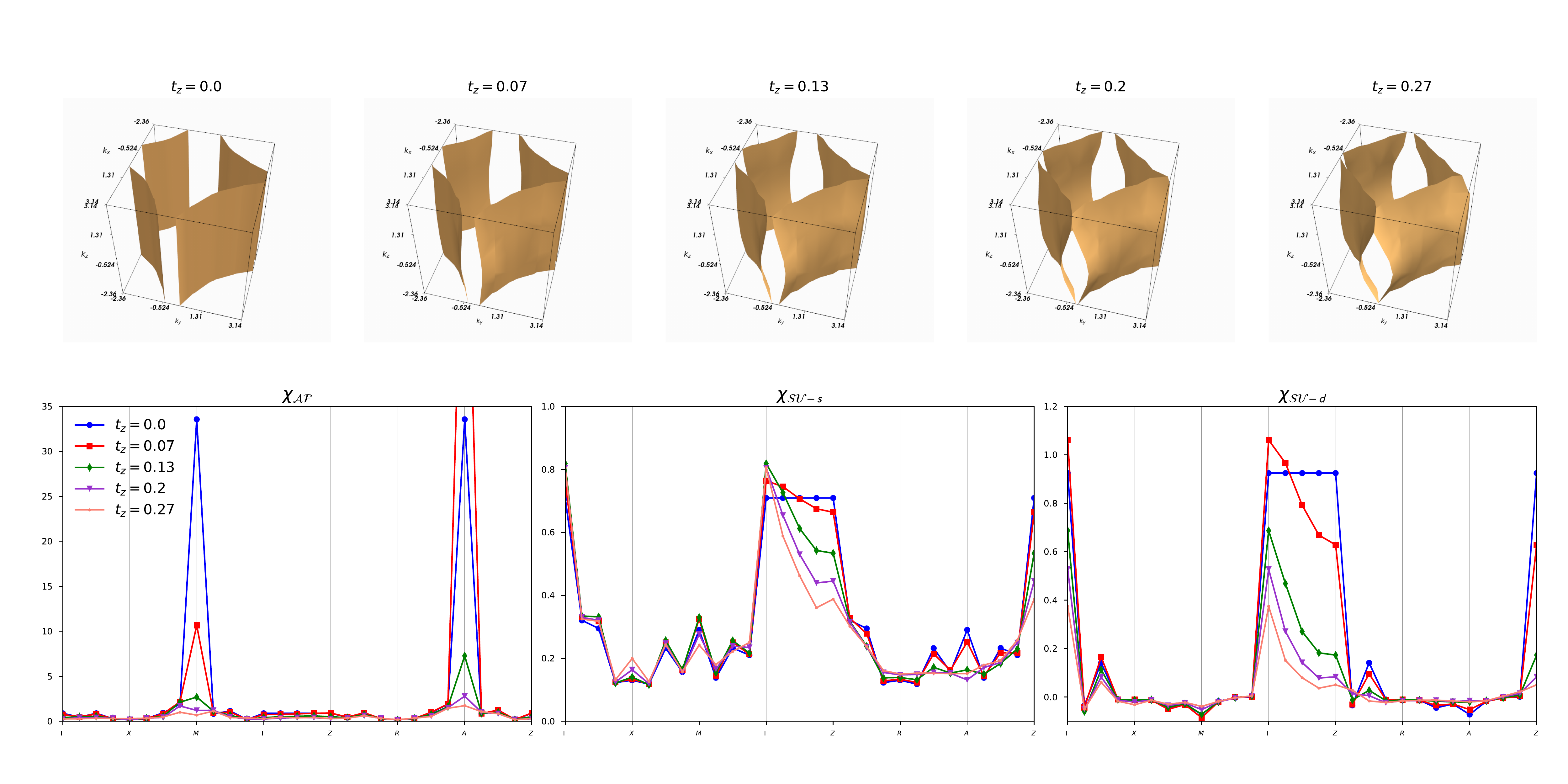}
\caption{The Fermi surface (top), the antiferromagnetic and superconducting susceptibilities ($s$, $d_{x^2-y^2}$) of the quasi-2D Hubbard model for various values of the out of plane hopping ($t_z$) at fixed Hubbard interaction ($U/W=0.25$) and fixed doping ($p=0.06$).}
\label{susPfs3DqHM}
\end{figure*}

Single particle properties appear stable across the series with Fig.\ref{nickSeriesFSsus} showing the differences in the Fermi surfaces and band structures for the model systems. Unlike changes to the AF and SU responses, the impact of hybridization is minimal at the single particle level. Overall, we find the effects of the hybridization with the interstitial-s driving novel physics, with the models showing nodeless $s,d_{z^2}$-type superconductivity dominating the reduced AF and d-superconducting fluctuations. The differences in the impact of the two bands can partially be attributed to the relative strengths of the hybridization but our studies indicate a large degree of hybridization ($V_d$) with the ${d_z^2}$ for the materials considered would be insufficient as the mechanism for the suppression of antiferromagnetic spin response in the Nickelates.

\subsection{Application of the Hubbard Model to the Nickelates}
\label{HubbardModel}

The primary model previously utilized in modeling superconductivity in the Nickelates is the single-band Hubbard model on a two dimensional square lattice with a dispersion given by $\xi_k =-2t(\cos(k_x)+\cos(k_y))+4t^\prime\cos(k_x)\cos(k_y)-2t^{\prime\prime}(\cos(2k_x)+\cos(2k_y))$. Discarding the roles of the hybridization ($V$) and out of plane hopping ($t_z$) could be a valid approximation for some parameter regimes,  so the results for the 2D Hubbard model can serve as a baseline from which we can measure the impact of accounting for those additional terms. From the {\it{ab-initio}} values for the $RNiO_2$ series the hopping in the Ni orbital is $t_{Ni}=0.375eV$ with the remaining hopping parameters going roughly as $t^\prime=0.25t$ and $t^{\prime\prime}=0.125t$\cite{been2021electronic}. The local Hubbard interaction of the Nd-Nickelate calculated via cRPA is found to be $U=7t-8t$ but the restriction of the fRG to moderate coupling limits our studies to $U\leq 4t$. We begin this section by analyzing the response of the single band Nickelate model to changes in the Hubbard interaction. The Hubbard model is given by
\begin{align}
   \mathcal{H}&=\sum_{k\sigma}\xi^{Ni}_kf_{k\sigma}^\dagger f_{k\sigma} +U\sum_in_{i\uparrow}^fn_{i\downarrow}^f-\mu\sum_{i\sigma}n_{i\sigma}^f
\end{align}
where $\xi^{Ni}$ is the dispersion given above. The spin and superconducting response of the model for various values of the Hubbard coupling is shown in Fig.\ref{afNick2DUdep}. As expected, increasing the coupling leads to enhanced antiferromagnetic and $d$-superconducting correlations across the doping range while ferromagnetic and $s$-superconducting fluctuations remain unaffected. If this trend persists into the strong coupling regime we have the familiar picture of antiferromagnetic spin fluctuations driving $d_{x^2-y^2}$ superconducting order.

\begin{figure}
\centering
\includegraphics[scale=0.29]{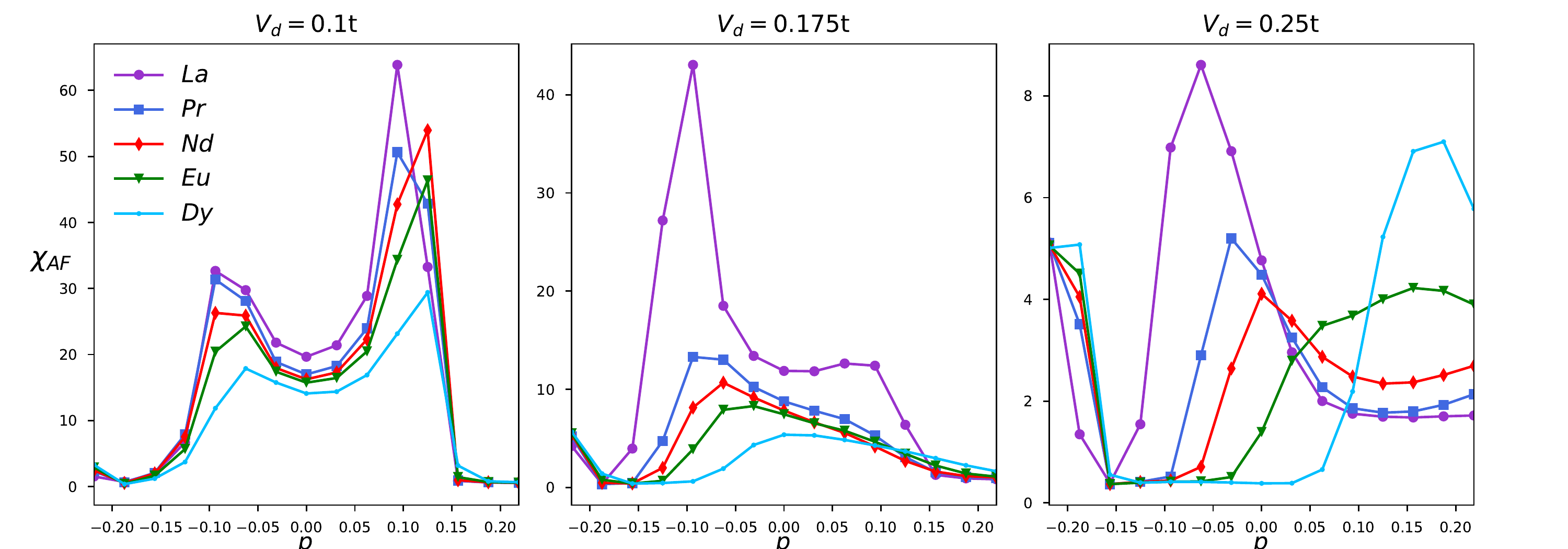}
\includegraphics[scale=0.29]{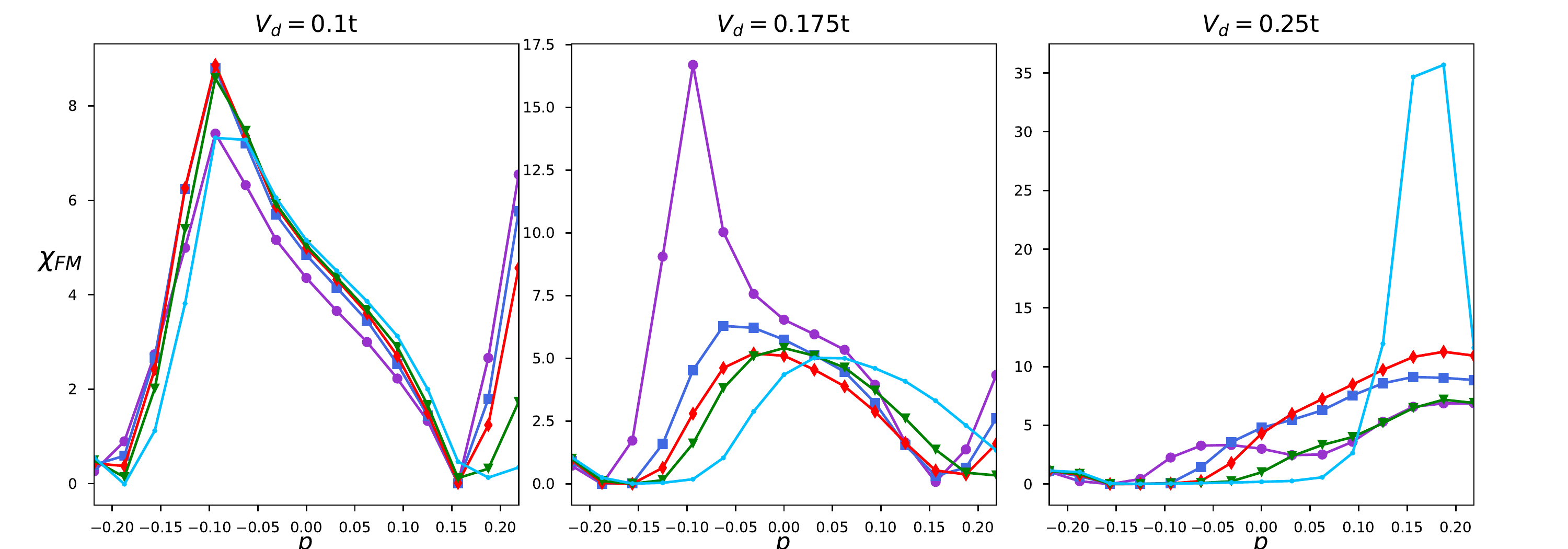}
\includegraphics[scale=0.29]{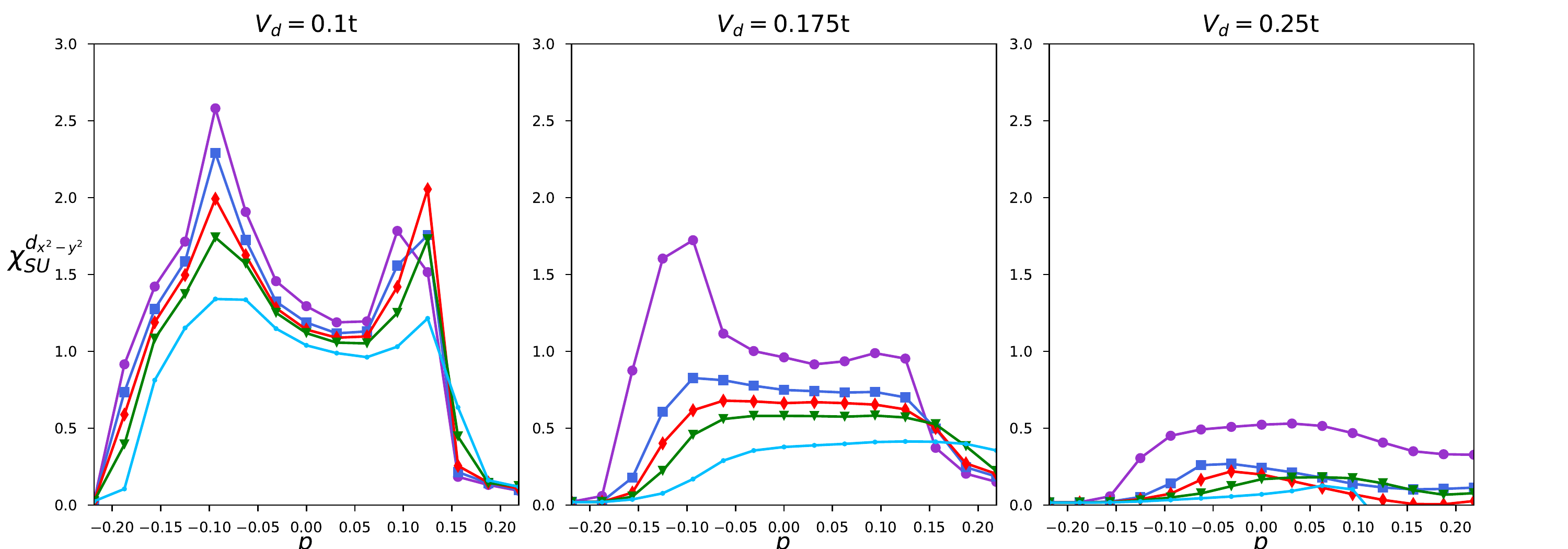}
\includegraphics[scale=0.29]{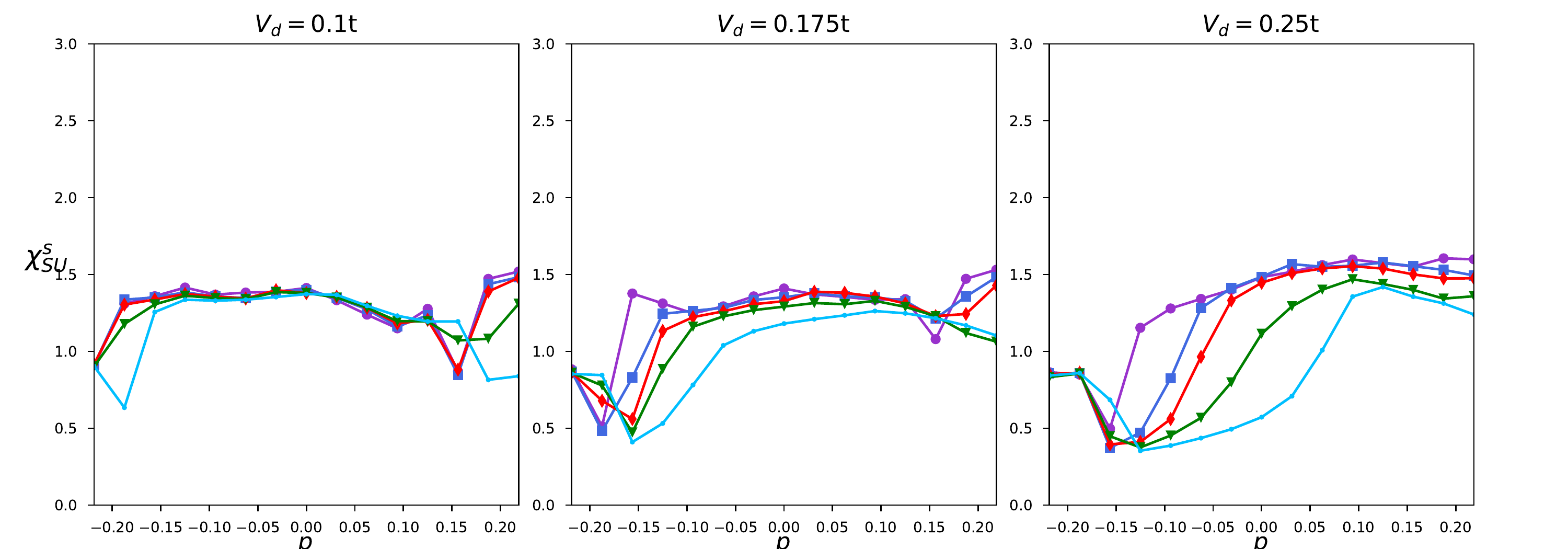}
\caption{The antiferromagnetic(top), ferromagnetic (middle) $d_{x^2-y^2}$ and s-superconducting susceptibilities as a function of doping for the 2D Nickelate model for different levels of hybridization with the series of $R$-bands with fixed Hubbard interaction ($U=3t$).}
\label{susResp2DR}
\end{figure}

Experimental findings of nodeless superconductivity in the Nd-Nickelates have led to a proposal of a 2D to 3D dimensional crossover, facilitated by orbital hybridization with the rare earth via the interstitial-s orbital\cite{chow2022pairing,chow2023dimensionality}. Previous observations of such crossovers seen in Ti$_2$Mo$_6$Se$_6$, where interchain hopping drives a 1D$\rightarrow$3D superconducting crossover and pressure induces a 2D$\rightarrow$3D crossover in the Bi2212 Cuprate superconductors suggest weak inter-dimensional coupling as the likely candidate for the crossover\cite{mitra2018dimensional,zhou2022quantum,guo2020crossover}.  We can analyze the impact of out of plane hopping on the Hamiltonian above by introducing an additional $-2t_z\cos(k_z)$ into the dispersion relation, $\xi_k$. We compensate for the increase in bandwidth by increasing the interaction strength with $t_z$ and fix $U/W$. The changes to the Fermi surface as well as the momentum dependence of spin and superconducting fluctuations for the hole doped case are shown in Fig.\ref{susPfs3DqHM}. Throughout the doping range we observe a drop in the spin response at $(\pi,\pi,0)$ as soon as we allow for out-of-plane hopping ($t_z$). This drop is accompanied by enhanced superconducting fluctuations in the $s$ and $d$ channels. The initial enhancement is smaller in the $s$ channel but as we further increase $t_z$ we begin to see the suppression of the planar $d_{x^2-y^2}$ response leaving the $s-wave$ as the dominant superconducting response. Moving beyond weak coupling the response of the model between the 2D ($t_z=0$) and 3D ($t_z=1$) limits is shown in Fig.\ref{susNick3DzUdep}. Although the inclusion of $t_z$ suppresses the spin response of the system, we still see a strong $(\pi,\pi,\pi)$-antiferromagnetic response and an in-plane spin response at values of $t_z$ of relevance to the Nickelates ($t_z<0.15t_{x,y}$). Extrapolating our results to the strongly coupled regime, we can expect the initial enhancement in spin and $d_{x^2-y^2}$-superconducting response to be more pronounced, which combined with the decrease in the $s$-SU response observed with increasing $U$ (Fig.\ref{afNick2DUdep}) should push the transition from $d\rightarrow s$ occurring at $t_z\sim0.15t_{x,y}$ for $U=0.3W$ further from the $t_z$ value for the Nickelates. These results suggest out of plane hopping alone as being insufficient to capture the crossover observed in the Nickelates.

\begin{figure}
\centering
\includegraphics[scale=0.29]{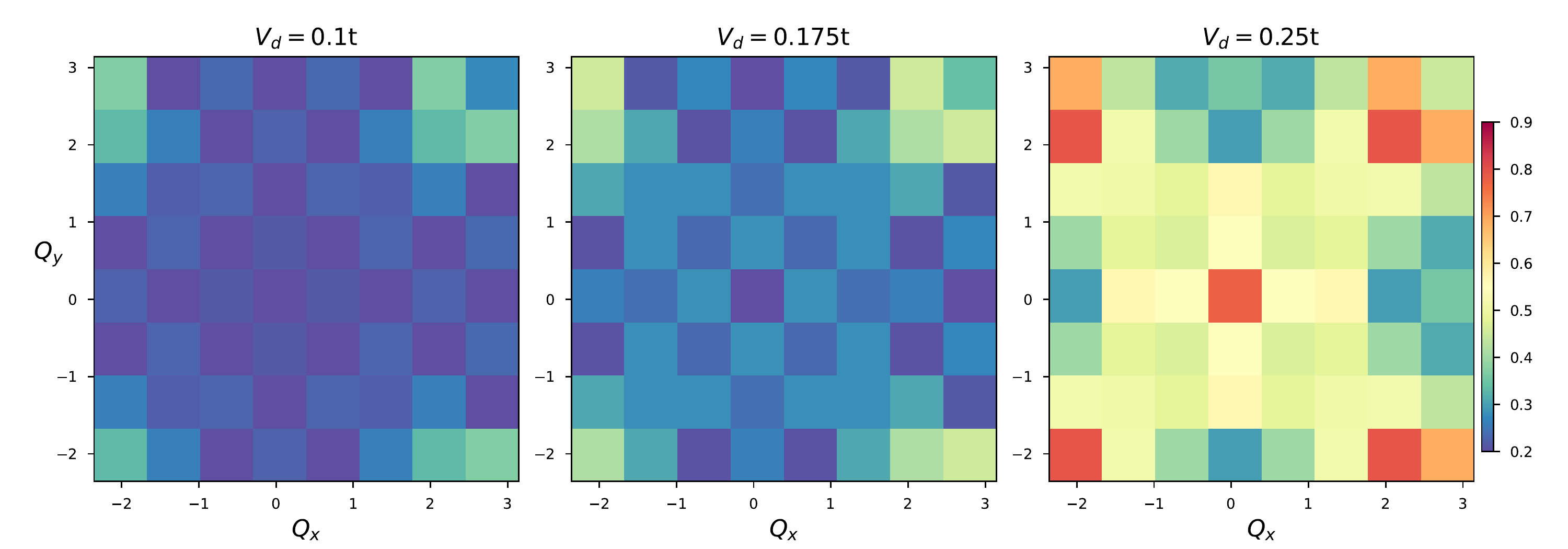}
\includegraphics[scale=0.29]{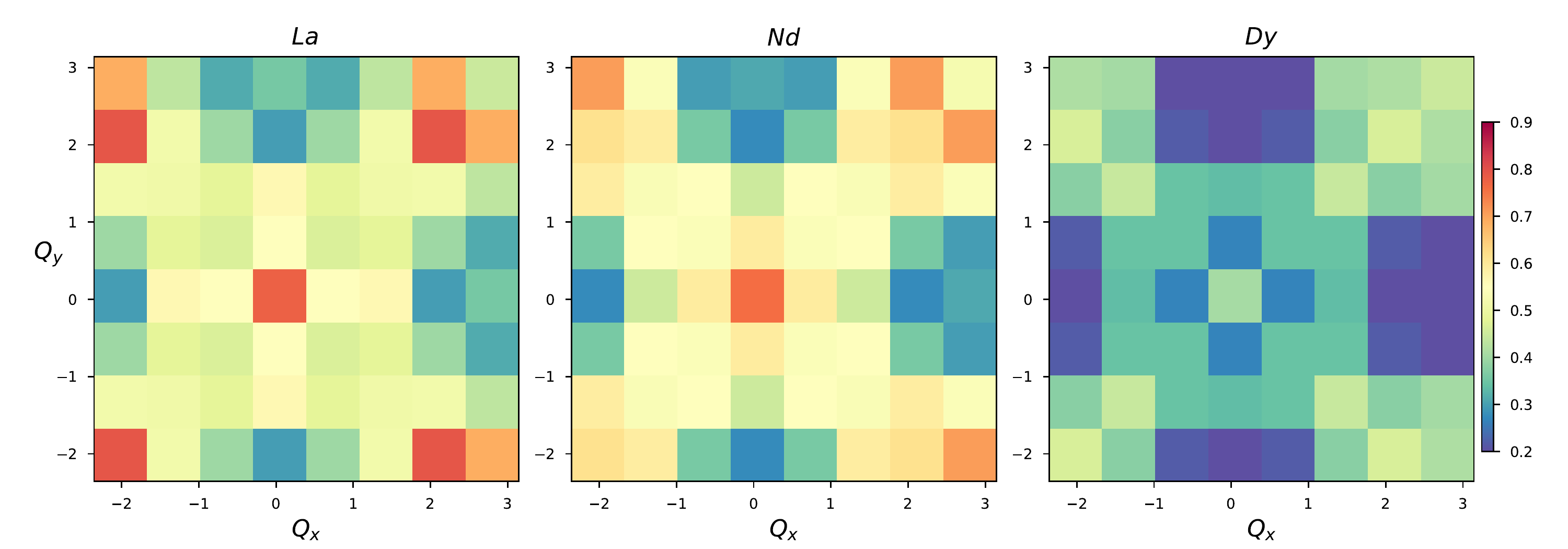}
\caption{Momentum dependence of the charge response for the 2D Nickelate model hybridized with a  series of $R$ bands at fixed Hubbard coupling. The charge response for different values of hybridization (top panels) for the La system and the charge response across $R$ at fixed hybridization are shown.}
\label{chargeResp2DVdR}
\end{figure}

\begin{figure*}
\centering
\includegraphics[scale=0.4]{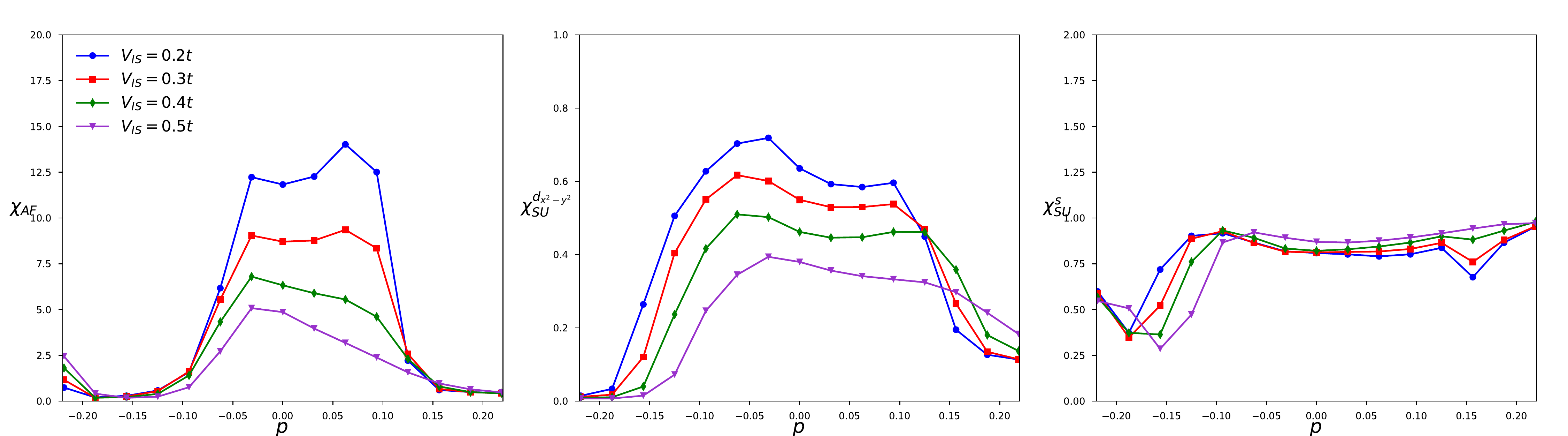}
\caption{The antiferromagnetic, $d_{x^2-y^2}$ and $s$ superconducting susceptibilities as a function of doping for the 2D Nickelate model hybridized with the interstitial-s orbital at fixed Hubbard interaction ($U=0.3W$). }
\label{susResp2DIS}
\end{figure*}

\subsection{Generalized Periodic Anderson Model}
\label{gPAM}
The impact of hybridization with the 3D-R and interstitial-s bands on the 2D NiO$_2$ layers can be explored at the two dimensional level as the non-interacting conducting bands can be integrated out. Summing over the three dimensional component of the the hybridization potential given in Eq.\ref{hybPot}, we have
\begin{align}
    \Delta_{2D}(\omega,k)=\frac{1}{N}\sum_{k_z}\frac{V_k^2}{i\omega-\xi_k^c}
\end{align}
where $V_k$ and $\xi_c$ correspond to the hybridization and dispersion of the R and interstitial-s bands. In the previous sections, the value of the hybridization was fixed to the {\it ab-initio} values derived in previous works \cite{gu2020substantial,been2021electronic}; here we analyze the impact of varying the hybridization strength at different doping levels on the response of the system. Though the exact range of values relevant to the Nickelates is yet to be determined, our previous results had the hybridization with interstitial-s an order of magnitude stronger than the R-band hybridization calculated for the {\it ab-initio} models. Keeping to these values, we separately explore changes to the 2D NiO$_2$ response as the R-hybridization is varied up to $V_d\sim 0.25t$ and up to $V_{S}\sim t$ for the interstitial $s$-band.

The response of the system to increasing $V_d$ hybridization for the various R bands is shown in Fig.\ref{susResp2DR}. The general trend seen across the R bands as the hybridization is increased is the reduction of the antiferromagnetic and $d_{x^2-y^2}$-SU response. Throughout the doping range considered the AF response is suppressed with a more significant reduction in the hole doped regime as $V_d$ is increased. The degree of suppression of the two orders is matched, although we note that the remnants of the d-SU response in the hole doped regime may still be sufficient to drive a superconducting transition at a lower temperature. In general, as the hybridization is varied the response is similar for the various R bands considered with the La and Dy bands breaking away from the trends in certain doping ranges. The uniform decrease in response seen across the R bands (particularly in the weak V regime) is directly due to the increase in the separation ($\epsilon_0^R$) from the Ni band. The bandwidth ($t_R^{[1,0,1]}$) is also smallest for the La band which explains the relative ordering of the superconducting response as the hybridization is increased. Similarly, the larger bandwidth of the Dy band appears to lead to an enhanced ferromagnetic order in the hole doped regime. At weak hybridization the s-type superconducting response is essentially degenerate for the R-bands across the doping regime, but as V is increased we see weak suppression in the $p<0$ regime which is likely driven by the out of plane hopping ($t_R^{[0,0,2]}$) which increases monotonically across the R series. Larger hybridization with the $R$ bands also enhances incommensurate charge ordering in the hole doped regime. Charge ordering in the Nickelates is well documented, although it is absent in our results at weak hybridization presented in the previous section. The effects of increasing hybridization with $R$ bands for the La system is shown in Fig.\ref{chargeResp2DVdR}. Although the response varies across the series we see hybridization in the system as enhancing metallic and charge ordered states in competition with superconductivity.

\begin{figure*}
\centering
\includegraphics[scale=0.4]{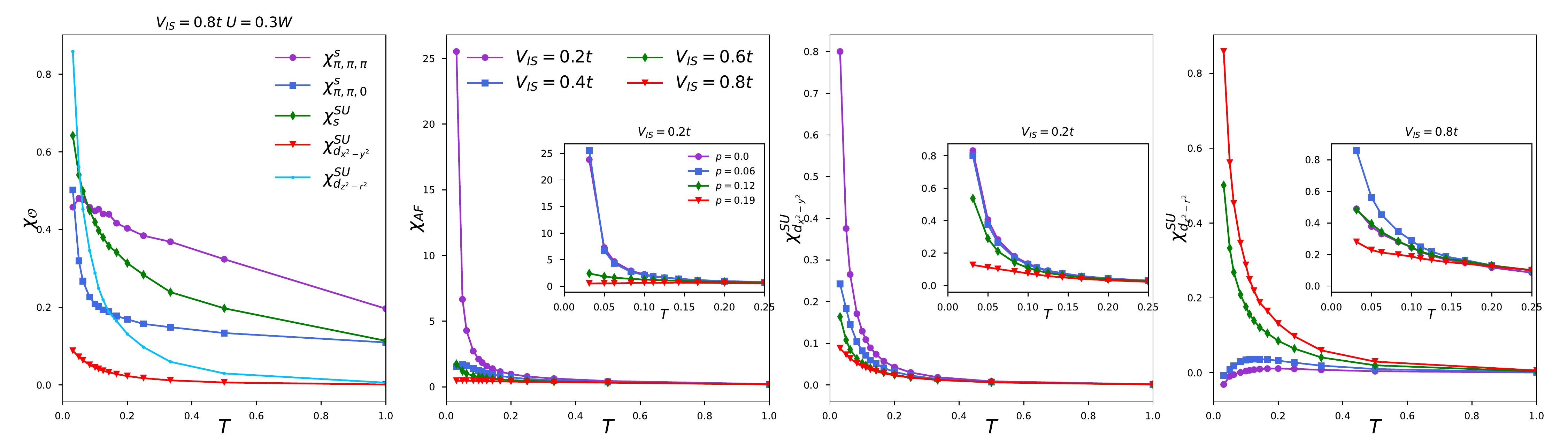}
\caption{Temperature dependence of the spin and superconducting fluctuations of the 2D Nickelate model for different values of hybridization with the interstitial-$s$ band at fixed Hubbard interaction ($U=0.3W$). The inset shows the doping dependence of the orders at their peak.}
\label{susResp3DIS}
\end{figure*}

\begin{figure}
\centering
\includegraphics[scale=0.6]{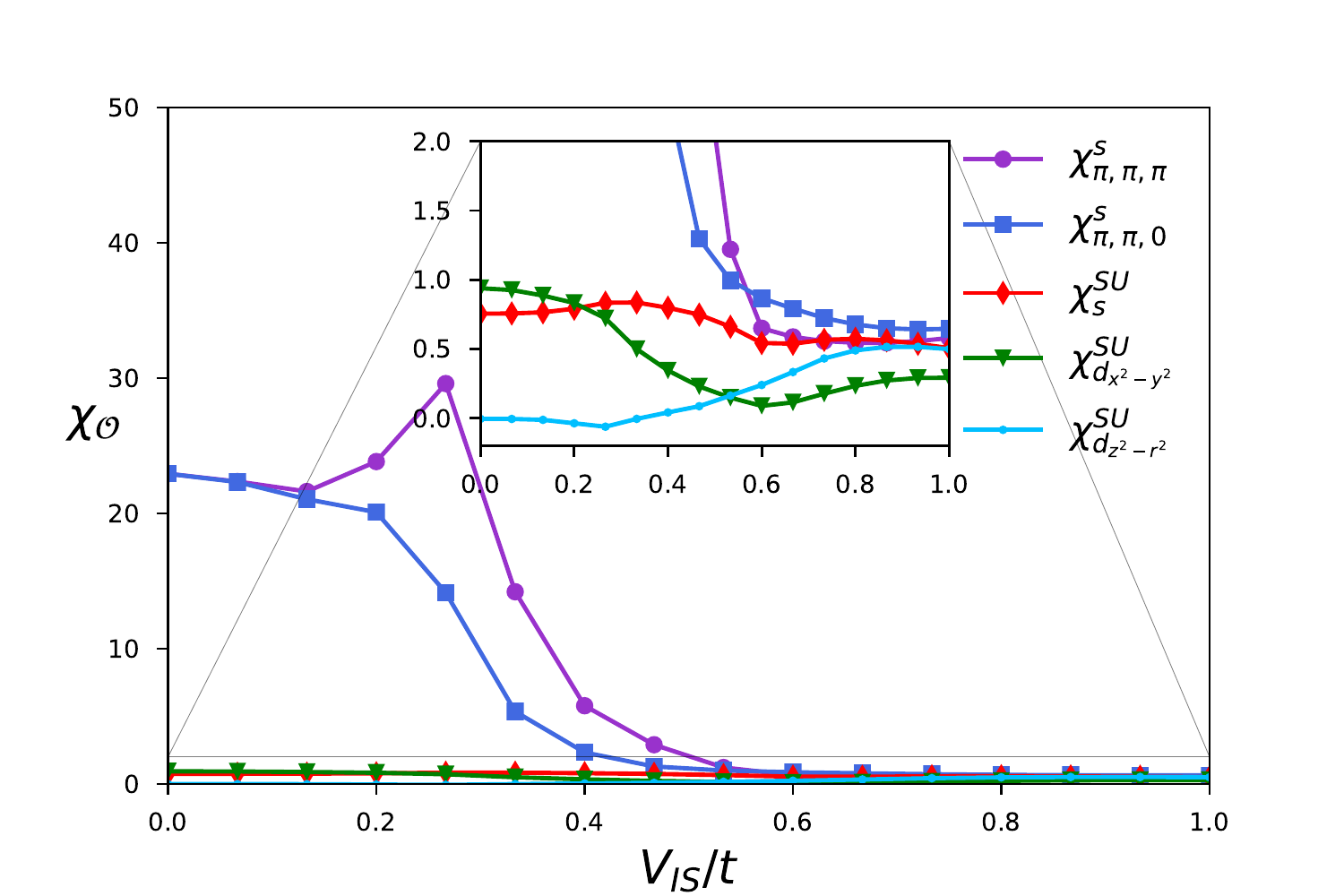}
\caption{The spin and superconducting susceptibilities of the 2D Hubbard model ($U=0.3W$) of the Nickelates hybridizing with the 3D interstitial-s band as a function of the hybridization at fixed doping ($p=0$).}
\label{susNick3DVISdep}
\end{figure}

A similar suppression of antiferromagnetic and d-SU is seen for hybridization with the interstitial s band shown in Fig.\ref{susResp2DIS}. Unlike the results for the $d_z$ band, large hybridization to the interstitial-s suppresses the antiferromagnetic response only in the hole doped region. The rate of suppression appears much higher for the AF response than the d-SU response though a complete suppression as seen in the Nickelate compounds up to 2K requires $V_s$ in excess of $\sim 0.5t$. This suppression is accompanied by the slight enhancement of s-type superconductivity suggesting a dominant $V_s$ is likely to lead to nodeless superconductivity in the Nickelates.

The role of hybridization with the 3D interstitial-s band in driving a 2D-3D superconducting crossover in the nickelates can be analyzed by coupling the band to the 2D Hubbard model of the Nickleates studied in the previous section. Analysis of the crossover requires retaining the $k_z$ dependence of the band, but we start with an initial $k_z$ independent Nickelate Hubbard model. At moderate coupling ($U=0.3W$) the Hubbard model shows strong AF order and $d-SU$ correlations across the doping range. The changes to the system's response due to hybridization with the interstitial-s band are shown in Fig.\ref{susNick3DVISdep}. With increasing hybridization, the antiferromagnetic and $d-SU$ responses are suppressed and a 3D $d_{z^2-r^2}$ superconducting response emerges in the same hole doping range ($p\sim 0.09$) that showed a strong $d-SU$ response. The evolution of the system's response at large hybridization as a function of temperature is shown in the left panel of the Fig.\ref{susResp3DIS}. Both the $s$ and the $d_z^2$ superconducting responses appear robust at low temperatures,  although given the suppression of the $s$-SU response at moderate Hubbard coupling we have shown in Fig.\ref{afNick2DUdep}, we expect the $d_{z^2}$-SU order to be dominant at strong coupling. This is further supported by the small change seen in the $s$-SU response for different values of interstitial hybridization shown in Fig.\ref{susNick3DVISdep}. Returning to the Nickelate compounds approximated by these models, the magnitude of hybridization appears to explain both the absence of spin fluctuations and the presence of charge order in the system.

\section{Conclusions}
\label{summary}
We have utilized the decoupled fRG to study separately the contributions from momentum dependent hybridization and out of plane hopping to the correlated $3d_{x^2-y^2}$-Ni electron band and their role in determining the details of the superconducting order observed in the system. We began by cataloging the impact of hybridization and frustration at a variety of doping levels and temperatures in Periodic Anderson Hamiltonians (realized in both two and three dimensions). Beyond serving as a test of our methods, we show PAM systems can stabilize a variety of phases, with the 2D model showing regions of anti-ferromagnetism, ferromagnetism, and s and $d_{x^2-y^2}$-superconductivity while the hybridization in the 3D system enhances the extended-s and a $d_{z^2-r^2}$ superconducting order in addition to the local orders observed in the 2D system. Frustration due to the momentum dependence of the hybridization drives competition between ferromagnetic and antiferromagnetic orders with superconducting fluctuations tucked in at the transition between the spin orders. We find frustrated hybridization with the local and nearest neighbor enhancing d-type superconductivity in 2D while competition between local hybridization and an interstitial-$s$ orbital drives $d_{z^2-r^2}$ superconductivity in the 3D PAM .

Addressing the role of hybridization in the Nickelates requires specifying the dispersing and conducting bands in the 3D generalized PAM model. Our study utilized a series of previously proposed {\it ab-initio} dispersing PAM models of the Nickelates\cite{been2021electronic}. The fRG flow for these models captures a doping dependent d-type superconductivity accompanied by a strong AF response across the series. Electronic correlations arising from the Hubbard coupling drive a strong antiferromagnetic spin response which due to the quasi-2D nature of the Nickelates peaks at A ($(\pi,\pi,\pi)$) and M ($(\pi,\pi,0)$). These spin fluctuations mediate pairing with stronger coupling enhancing the calculated AF and $d-SU$ response of the system. The strong AF is incongruent with experimental results and additional hybridization with the interstitial-s orbital is required to suppress the antiferromagnetic correlations. However, we found the additional hybridization to also suppress the $d$-superconducting response while the $s$ superconducting response remains unaffected over a wide range of hybridizations. The hybridization does enhance the extended d$_{z^2-r^2}$ superconducting response suggesting nodeless superconductivity as the norm in the Nickelates. Of the series of rare earth atoms we considered, we found, in the absence of coupling to the interstitial-s, the $d_{x^2-y^2}$  superconducting response peaking in the Eu and Pr systems which at the level of band structure appears due to the quasi-2D nature of the models.

Finally, guided by recent experimental studies of the Nickelates that suggest the initial d-type superconductivity in the $\mathrm{NiO_2}$ planes as a precursor that initiates a nodeless 3D superconducting state driven by the hybridization, we considered mechanisms for the superconducting crossover. To this end, we separately analyzed the effects of weak out of plane hopping ($t_z$) and hybridization with the interstitial-s band ($V_{IS}$) on a 2D Nickelate model with $d_{x^2-y^2}$ superconducting correlations. With the Hubbard coupling ($U/W$) fixed, we found that weak values of $t_z$ ($<0.2t$) work to the detriment of the AF and d-superconducting order while inducing little change in the $s-SU$ response. The effects of hybridization are more pronounced and leads to a dramatic suppression of antiferromagnetic correlations in the hole doping region of interest with hybridization to the interstitial-s having the additional effect of enhancing s and $d_{z^2}$-superconducting order in the same region. These results suggest the strength of hybridization to the interstitial-s as a control for the transition between 2D and 3D superconducting order. Given the overall negative effect of electronic correlations on the $s$-SU, the extended $d_{z^2-r^2}$ order is the likely candidate for nodeless superconductivity in the Nickelates. A full resolution of the Nickelate picture requires further studies that address the impact of ordering in the $R$ spacer layer and their interplay with strong hybridization ($V$) in the Nickelates. Work in these directions is currently underway.

\section{Acknowledgements}
We wish to thank Ka-Ming Tam and Shan-Wen Tsai for illuminating discussions during the course of this work. We would also like to thank Boston University's Research Computing Services for their technical support and computational resources.
\bibliography{generalPeriodicAndNick.bib}
\appendix

\end{document}